\documentclass[11pt,a4paper]{article}
\usepackage{amsmath,amssymb,amsthm,graphicx,braket,multirow,authblk,amsthm,dcolumn,latexsym,subcaption,mathtools,cite,cancel}
\usepackage[normalem]{ulem}
\usepackage[a4paper]{geometry}
\DeclareSymbolFontAlphabet{\amsmathbb}{AMSb}%
\usepackage[most]{tcolorbox}
\usepackage[T1]{fontenc}
\usepackage[utf8]{inputenc}
\usepackage{graphicx}
\usepackage[unicode=true,bookmarks=true,bookmarksopen=false,breaklinks=false,pdfborder={0 0 0},colorlinks=true]{hyperref}

\usepackage{nicematrix, tikz}
\usetikzlibrary{calc}
\usepackage{xcolor}
\usepackage{blkarray}
\usepackage{placeins}
\definecolor{cblue}{rgb}{0.16, 0.32, 0.75}
\definecolor{cred}{rgb}{0.7, 0.11, 0.11}
\hypersetup{%
	,linkcolor=cred
	,citecolor=cblue
	,urlcolor=blue
}
\def\<{\langle}
\def\>{\rangle}

\newcommand{\Tr}{\mathrm{Tr}}
\newtheorem{Example}{Example}
\newtheorem{theorem}{Theorem}%[section]
\newtheorem{proposition}{Proposition}%[section]
\newtheorem{corollary}{Corollary}%[section]
\newtheorem{definition}{Definition}%[section]
\newtheorem{remark}{Remark}
\date{}
\def\oper{{\mathchoice{\rm 1\mskip-4mu l}{\rm 1\mskip-4mu l}
		{\rm 1\mskip-4.5mu l}{\rm 1\mskip-5mu l}}}

\newcommand{\ketbra}[2]{| #1 \rangle\!\langle #2 | }

\newcommand{\BH}{\mathcal{B}(\mathcal{H})}

\newcommand{\LA}{\mathcal{L}^A}
\newcommand{\LB}{\mathcal{L}^B}
\newcommand{\LC}{\mathcal{L}^C}
\newcommand{\BAa}{\mathcal{B}(\mathcal{H}^{A_0})}
\newcommand{\BAb}{\mathcal{B}(\mathcal{H}^{A_1})}
\newcommand{\BAA}{\mathcal{B}(\mathcal{H}^{A_0A_1})}
\newcommand{\BBa}{\mathcal{B}(\mathcal{H}^{B_0})}
\newcommand{\BBb}{\mathcal{B}(\mathcal{H}^{B_1})}
\newcommand{\BBB}{\mathcal{B}(\mathcal{H}^{B_0B_1})}

\newcommand{\HAa}{\mathcal{H}^{A_0}}
\newcommand{\HAb}{\mathcal{H}^{A_1}}
\newcommand{\HBa}{\mathcal{H}^{B_0}}
\newcommand{\HBb}{\mathcal{H}^{B_1}}
\newcommand{\Md}{\mathcal{M}_d}

\DeclareMathAlphabet\mathbfcal{OMS}{cmsy}{b}{n}

\date{}

\begin{document}	

\title{\textbf{Diagonal Unitary Covariant Superchannels}}

	\author[1]{Dariusz Chru\'sci\'nski\footnote{darch@fizyka.umk.pl}}
	\affil[$1$]{\small Institute of Physics, Faculty of Physics, Astronomy and Informatics, Nicolaus Copernicus University, Grudziądzka 5/7, 87-100 Toru\'n, Poland}
    \author[1,2]{Vivek Pandey\footnote{vivekpandey3881@gmail.com}}
	\affil[$2$]{\small Tata Institute of Fundamental Research Hyderabad, 36/P, Gopanpally Village,
Serilingampally Mandal, Hyderabad, Telangana 500046, India}
    \author[3,4]{Sohail\footnote{sohail@chapman.edu}}
    \affil[$3$]{\small Institute for Quantum Studies, Chapman University, Orange, CA 92866, USA}
	\affil[$4$]{\small Department of Computer Science, Texas Tech University, Lubbock, TX 79409, USA}
	\maketitle
	\vspace{-0.5cm}	
	
	\begin{abstract}
	We present a complete characterization of diagonal unitary covariant (DU-covariant) superchannels, i.e. higher-order transformations transforming quantum channels into themselves.  Necessary and sufficient conditions for complete positivity and trace preservation are derived and the canonical decomposition describing DU-covariant superchannels is provided. The presented framework unifies and extends known families of covariant quantum channels and enables explicit analysis of their action on physically relevant examples, including amplitude-damping, bit-flip, and Pauli channels. Our results provide a practical toolbox for symmetry-restricted  higher-order quantum processes and offer a setting for exploring open problems such as the celebrated PPT$^2$ conjecture.
	\end{abstract}

\section{Introduction}

Symmetry plays a prominent role in quantum physics, providing both conceptual insight and powerful technical tools for simplifying the analysis of complex systems. In quantum information theory \cite{QIT,Wilde2013} invariance under local unitary operations has been particularly important in the analysis of quantum states of composite systems. Key examples of unitary invariant states include celebrated isotropic and Werner states, which play a fundamental role in various quantum information problems \cite{Werner1989, Horodecki1999,HHHH, VollbrechtWerner2001, Chruscinski2006a, Chruscinski2006b}. {At} the level of quantum channels, covariance under group actions leads to strong structural constraints, exact characterizations, and clear operational interpretations \cite{Scutaru1979,Holevo1993,Holevo1996}.  The
prominent examples of covariant channels include unitary-covariant depolarizing channels, transpose depolarizing
channels, phase-covariant, and diagonal unitary covariant channels \cite{Singh_2021,Nechita_2021,Singh2022,Nechita_2025}.

Consider  a group $G$ and its two unitary representations  in $\HAa$ and $\HAb$, respectively. We say that a channel $\Phi : \BAa \to \BAb$ is $G$-covariant w.r.t. these representations  if for all $g \in G$,

\begin{equation}   \label{}
    \Phi \circ \mathcal{U}_g = \mathcal{V}_g \circ \Phi ,
\end{equation}
where
 $ \mathcal{U}_g(X) := U_g X U_g^\dagger$ and $\mathcal{V}_g(X) := V_g X V_g^\dagger$ \cite{Holevo2011}.
  
Channels  covariant with respect to the irreducible unitary representation are called
irreducibly covariant. It turns out that for such channels Holevo capacity is additive and proportional to the
minimum output entropy~\cite{holevo2002,Jozsa1996,Datta2005, Mozrzymas_2017}. This discovery started a series of articles dealing with additivity and
multiplicity of minimum output entropy for irreducibly covariant quantum channels~\cite{Werner2002, Fannes2004, Holevo:2005,  Fukuda2006, Fukuda2017, Mozrzymas_2017}. Covariant quantum channels play an important role in several areas of quantum information theory such as quantum error correction\cite{Faist_2020,Zhou_2021,Hayden2021}, programmability \cite{Gschwendtner_2021}, entanglement theory~\cite{Christandl2019,bauml2019,Bardet_2020,Singh_2021,Nechita_2025} and quantum Shannon theory~\cite{Werner2002,Holevo:2005,holevo2002,datta2004,Pirandola2011,Pirandola2017,Wilde_2017,Das2019,Das_2021,sohail2025}, etc. Moreover, their mathematical properties, which have been extensively studied for many years~\cite{AlNuwairan2014,Lee2022,Memarzadeh2022}, provide a more tractable computation of many information theoretic quantities defined for quantum channels, for example relative entropy of quantum channels~\cite{Gour2021, Felix2018, Yuan2019,sohail2025}, which have found operational meaning in several quantum information processing tasks.

Quantum superchannels provide a natural  generalization of quantum channels \cite{Chiribella, Giulio2008a, Zyczkowski2008}. 
Superchannels represent the most general physical maps transforming quantum channels into quantum channels and naturally arise in adaptive quantum protocols, channel discrimination, and programmable quantum dynamics\cite{Chiribella, Giulio2013, Gour_2019, Gour2021}. 
The theory of quantum superchannels has emerged as a powerful tool to tackle several challenging problems in quantum information theory\cite{Giulio2008,Giulio2008b,Ziman2008,Bisio2009,Giulio2009,Giulio2009a,Bisio2010,Giulio2013}.
Quantum superchannels also play an important role in the dynamical quantum resource theories~\cite{Wang_2023,Diaz2018,Gour2021a}, the recoverability  and measurement of quantum channels~\cite{Burniston2020,sohail2025}. 

{ Quantum combs constitute the general framework for higher-order quantum maps describing physically admissible transformations of quantum operations~\cite{Chiribella, Giulio2009a, Pollock2018, Pollock2018NonMarkovian, Milz2021Quantum, Jencova2024Structure, Milz2024Characterising}. A superchannel is precisely a one-slot quantum comb, i.e. the most general physical transformation mapping quantum channels to quantum channels. In this work we characterize the subclass of one-slot quantum combs satisfying diagonal unitary covariance. Quantum combs are mathematically equivalent to process tensors, which provide the standard description of non-Markovian quantum stochastic processes. A superchannel is the simplest one-slot process tensor, whereas multi-slot process tensors describe general memory-bearing quantum evolutions with intermediate interventions. Thus, the diagonal unitary covariant superchannels studied here may be viewed as the elementary building blocks of symmetry-constrained non-Markovian quantum processes \cite{Pollock2018,Pollock2018NonMarkovian}.}

While the general structure of superchannels is well understood (cf. the review \cite{Gour_2019}), their explicit characterization remains challenging due to the high dimensionality of the space and the nontrivial constraints imposed by complete positivity and trace preservation. As in the case of quantum channels, imposing symmetry constraints on superchannels offers a route toward tractable classifications.  

{Symmetry constraints on higher-order quantum maps arise naturally when manipulating families of symmetric channels. Examples include covariant channel simulation, programmable quantum processors respecting a symmetry group, and resource theories in which free superchannels are required to preserve the symmetry of channels. The present work provides a complete characterization of superchannels covariant with respect to diagonal unitary and diagonal orthogonal groups.}

We show that imposing covariance with respect to diagonal unitary actions at both the input and output levels therefore yields a physically motivated and mathematically interesting class of superchannels. Diagonal unitary covariance leads to a canonical decomposition of superchannels into independent components acting on diagonal and off-diagonal sectors of Choi matrices corresponding to quantum channels. This decomposition generalizes the known results for diagonal unitary covariant quantum channels \cite{Singh_2021} and provides a transparent interpretation of superchannel action in terms of classical superchannels combined with constrained transformations of coherence sectors. As an important application, we identify and analyze dephasing superchannels as a natural subclass of diagonal covariant superchannels. These superchannels map arbitrary quantum channels to dephasing channels and can be characterized by Schur multiplication at the level of the corresponding Choi matrices.  

%\paragraph{Outline of the paper.}
The paper is organized as follows.  
In Section~\ref{sec2}, we review the formalism of quantum superchannels and their representing maps. 
Section~\ref{sec3} introduces the notion of covariance for superchannels and provides several equivalent characterizations in terms of their representing maps and Choi matrices. In Section~\ref{sec4}, we focus on unitary covariant superchannels and derive their general structure. Both the unitary covariant and conjugate unitary covariant cases are discussed.  

{  Sections~\ref{sec5} and~\ref{APP-ORT}  are devoted to diagonal unitary covariant and diagonal orthogonal covariant superchannels, respectively. Treating covariance as an intertwining condition, we exploit the decomposition of the corresponding operator spaces into weight (character) subspaces of the symmetry groups. This yields canonical decompositions of the representing maps and Choi matrices, together with complete characterizations of complete positivity and trace preservation.}

In Section~\ref{sec6}, we analyze the action of diagonal unitary covariant superchannels on quantum channels, showing how they affect diagonal and off--diagonal blocks independently.  
Section~\ref{sec7} is concerned with dephasing superchannels, which form an important subclass of diagonal unitary covariant superchannels and admit a natural interpretation in terms of Schur products and correlation matrices.
Section~\ref{sec8} presents several illustrative examples, including transformations of qubit amplitude--damping, bit--flip, and Pauli channels.  
In Section~\ref{sec9}, we introduce Pauli superchannels and show how they induce classical transformations between Pauli channels via bistochastic matrices.  
Finally, Section~\ref{sec10} contains concluding remarks and an outlook, including possible extensions of the framework and open problems. The technical details are relegated to a series of Appendices.

\section{Quantum Superchannels}\label{sec2} 

{Let $\BH$ denote the space of all operators acting
on a finite dimensional Hilbert space $\mathcal{H}$. In this paper we consider finite dimensional Hilbert spaces only.  $\BH$ is an inner product space equipped with the Hilbert-Schmidt
inner product $(X,Y)= \Tr(X^\dagger Y)$ for all $X, Y \in \BH$, where $X^\dag$ stands for the Hermitian conjugation. A linear map $\Phi :  \BAa \to \BAb$ is called to be positive if $\Phi(X) \geq 0$  for all positive semi-definite operators $X \in \BAa$. Denote by $\mathcal{M}_k$ an algebra of $k \times k$ complex matrices. A map $\Phi$ is $k$-positive if ${\rm id}_k \otimes \Phi$ is positive (${\rm id}_k$ stands for the identity map in $\mathcal{M}_k$) \cite{paulsen2002,book, Bhatia2006Positive, Watrous2018Theory, Wolf2012Quantum }. Finally, $\Phi$ is completely positive (CP) if it is $k$-positive for all $k=1,2,\ldots$. Completely positive trace-preserving map (CPTP) is called a quantum channel \cite{QIT}. Given $\Phi : \BAa \to \BAb$ one defines its adjoint $\Phi^\ddag : \BAb \to \BAa$  w.r.t. the Hilbert-Schmidt inner product 

\begin{equation}
    (\Phi(X),Y) =: (X,\Phi^\ddag(Y)) . 
\end{equation}
Recall, that $\Phi$ is CP if and only if $\Phi^\ddag$ is CP and $\Phi$ is trace-preserving if and only if $\Phi^{\ddag}$ is unital, i.e. $\Phi^{\ddag}\left(\oper_{A_1}\right) = \oper_{A_0}$ \cite{paulsen2002,book, Bhatia2006Positive, Watrous2018Theory, Wolf2012Quantum}. }

Consider now a vector space of linear {\em supermaps} 

\begin{equation}
    \Theta  : \LA  \to  \LB ,
\end{equation}
where
\begin{eqnarray}
    \mathcal{L}^A = \left\{ \Phi : \BAa \to \BAb \right\}  \ ,\ \ \ \mathcal{L}^B = \left\{ \Psi : \BBa \to \BBb \right\} ,
\end{eqnarray}
stands for spaces of linear maps. Denote by $d_{A_i}$ and $d_{B_i}$ corresponding dimensions of $\mathcal{H}^{A_i}$ and  $\mathcal{H}^{B_i}$ (for $i=0,1$). Let $\mathcal{L}^A_+$ and $\mathcal{L}^B_+$ denote a convex cone of completely positive  maps in $\LA$ and $\LB$, respectively. Following Refs.~\cite{Chiribella,Gour_2019}, we briefly review the basic concepts underlying quantum superchannels. 

\begin{definition} Consider a supermap $\Theta  : \LA  \to  \LB$:

\begin{enumerate}
    \item $\Theta$ is CP preserving if $\Theta[\Phi] \in \LB_+ $ for any $\Phi \in \LA_+$, i.e. $\Theta$ maps CP maps to CP maps,
    \item $\Theta$ is completely CP preserving if ${\rm Id}_C \otimes \Theta$ is CP preserving for any $C=(C_0,C_1)$ (where ${\rm Id}_C$ denotes an identity map in $\mathcal{L}^C$),
    \item $\Theta$ is TP preserving if $\Theta[\Phi]$ is TP for any TP map $\Phi \in \LA$.
\end{enumerate}
    
\end{definition}
Finally,

\begin{definition}
    A supermap $\Theta : \LA \to \LB$ is called a quantum superchannel if $\Theta$ is completely CP and TP preserving, i.e., it maps quantum channels from $\mathcal{L}^{CA}$ to quantum channels in $\mathcal{L}^{CB}$ {for all finite dimensional choice of $C=(C_0,C_1)$}. 
\end{definition}
Actually, any quantum superchannel can be realized as follows \cite{Chiribella,Gour_2019}

\begin{equation}
 \Theta[\Phi^{A_0 \rightarrow A_1}]   = \mathcal{M}^{RA_1 \rightarrow B_1}_{\rm{post}} \circ \left( {\rm id}^{R \to R} \otimes \Phi^{A_0 \rightarrow A_1}\right)  \circ \mathcal{M}^{B_0 \rightarrow RA_0}_{\rm{pre}}, \label{equ:superchannel_action}
  \end{equation}
  where $\mathcal{M}^{RA_1 \rightarrow B_1}_{\rm{post}}$ and $\mathcal{M}^{B_0 \rightarrow RA_0}_{\rm{pre}}$ are post-processing and pre-processing channels, respectively.

Any supermap $\Theta$  induces a linear map -- so called the representing map of $\Theta$

\begin{equation}
    \Delta_\Theta : \BAA \to \BBB ,
\end{equation}
defined via

\begin{equation}
    \Delta_\Theta := C_B \circ \Theta \circ C_A^{-1} ,
\end{equation}
where $C_A$ and $C_B$ denote the Choi-Jamio{\l}kowski isomorphism  \cite{CHOI1975285, Jamiolkowski1972Linear, dePillis1967Linear}

$$   C_A : \LA \to \BAA \ , \quad C_B : \LB \to \BBB . $$
Recall that fixing an arbitrary orthonormal basis  $\{e_1,\ldots,e_{d_{A0}}\}$ in $\HAa$ and introducing $e_{ij} := \ketbra{e_i}{e_j}$, we have \cite{CHOI1975285}

$$  C_A[\Phi] = \sum_{i,j=1}^{d_{A_0}} e_{ij} \otimes \Phi\left(e_{ij}\right)  .$$
{Interestingly, the Choi matrix was already analyzed in a seminal paper \cite{Sudarshan1961Stochastic} under the name $B$-matrix.}
Hence, $\Delta_\Theta$ realizes a supermap $\Theta$ on the level of Choi matrices, i.e.
if $C_\Phi$ denotes a Choi matrix of a map $\Phi \in \LA$, then

\begin{equation}
    \Delta_\Theta[C_\Phi] = C_{\Theta[\Phi]} .
\end{equation}
Moreover, given two supermaps $\Theta_1 : \LA \to \LB$ and $\Theta_2 : \LB \to \LC$, {the following composition rule holds:}

\begin{equation}
    \Delta_{\Theta_2 \circ \Theta_1} =  \Delta_{\Theta_2} \circ \Delta_{\Theta_1} .
\end{equation}

\begin{proposition} A supermap $\Theta$ is

\begin{enumerate}
    \item CP preserving iff $\Delta_\Theta$ is positive,
    \item completely CP preserving iff $\Delta_\Theta$ is completely positive,
    \item TP preserving iff there exists a linear map $\widetilde{\Delta}_\Theta : \BAa \to \BBa$ such that

\begin{equation}
    {\rm Tr}_{B_1} \circ \Delta_\Theta =  \widetilde{\Delta}_\Theta \circ  {\rm Tr}_{A_1} ,
\end{equation}
and  $\widetilde{\Delta}_\Theta(\oper_{A_0}) = \oper_{B_0}$.    
\end{enumerate}
    
\end{proposition}

Let us fix arbitrary orthonormal bases  $\{e_1,\ldots,e_{d_{A_0}}\}$ in $\HAa$ and 
$\{f_1,\ldots,f_{d_{A_1}}\}$ in $\HAb$. Introducing a set of maps $\epsilon_{ij;ab} : \BAa \to \BAb$ {given by}

\begin{equation}
    \epsilon_{ij;ab}\left(X\right) = \langle e_i|X| e_j \rangle  f_{ab} ,
\end{equation}
a  Choi matrix of a supermap $\Theta$ {can be defined as} \cite{Gour_2019}

\begin{equation}
    C_\Theta = \sum_{ij,ab} C_{\epsilon_{ij;ab}} \otimes C_{\Theta[\epsilon_{ij;ab}]} .
\end{equation}
Now, using $C_{\epsilon_{ij;ab}} = e_{ij} \otimes f_{ab} $, 
one obtains the following formula

\begin{equation}
    C_\Theta = \sum_{ij,ab} e_{ij} \otimes f_{ab} \otimes \Delta_\Theta\left(e_{ij} \otimes f_{ab}\right) ,
\end{equation}
that is, $C_\Theta =  C_{\Delta_\Theta}$, which shows that Choi matrices of $\Theta$ and its representing map $\Delta_\Theta$ coincide. Therefore, it is clear that $\Theta$ is completely CP preserving iff $C_\Theta \geq 0$. Recall that $C_\Phi$ is a Choi matrix of a quantum channel $\Phi \in \LA$ iff $C_\Phi \geq 0$ and ${\rm Tr}_{A_1} C_\Phi = \oper_{A_0}$. A similar property holds for superchannels as well.

\begin{proposition} An operator $C_\Theta$ defines a Choi matrix of a {superchannel} $\Theta : \LA \to \LB$ iff $C_\Theta \geq 0$ together with

\begin{equation}
    {\rm Tr}_{B_1} C_\Theta = C_0 \otimes \oper_{A_1} ,
\end{equation}
and ${\rm Tr}_{A_0} C_0 = \oper_{B_0}$. 
    
\end{proposition}

Observe that for a given quantum channel $\Phi \in \LA$,

\begin{equation}
    S_{ai} := {\rm Tr}\left( \Phi(e_{ii}) f_{aa}\right) ,
\end{equation}
defines a column stochastic matrix and provides a matrix representation of a classical channel $S : \mathbb{R}^{d_{A_0}} \to \mathbb{R}^{d_{A_1}}$. {Moreover, $S_{ai}$ can be rewritten as} 

\begin{equation}
    S_{ai} = \< e_i\otimes f_a| C_\Phi| e_i \otimes f_a \> ,
\end{equation}
where $C_\Phi$ stands for the Choi matrix of $\Phi$. Consider now a quantum superchannel $\Theta : \LA \to \LB$ and let $C_\Theta$ denote its Choi matrix. Define a matrix $T$  as

\begin{equation}
    T_{jb,ia} := \< e_i\otimes f_a; \tilde{e}_j \otimes \tilde{f}_b| C_\Theta | e_i \otimes f_a;\tilde{e}_j \otimes \tilde{f}_b \> , 
\end{equation}
where $\{\tilde{e}_j\}$ and $\{\tilde{f}_b\}$ denote orthonormal bases in $\mathcal{H}^{B_0}$ and $\mathcal{H}^{B_1}$, respectively. Hence $T_{jb,ia}$ is a $d_{A_0}d_{A_1}\times d_{B_0}d_{B_1}$ matrix with non-negative {entries} satisfying the following properties: {$\sum_{b=1}^{d_{B_1}}  T_{jb,ia}$ does not depend on $a$, that is, there exists a $t_{ji}$ matrix such that }

\begin{enumerate}
    \item $\sum_{b=1}^{d_{B_1}}  T_{jb,ia} = t_{ji}$, for all $a =1,\ldots,d_{A_1}$,
    \item $\sum_{i=1}^{d_{A_0}} t_{ji} = 1 $, for all $i=1,\ldots,d_{B_0}$.
\end{enumerate}
Note that $T_{jb,ia}$ represents a {\em classical superchannel} which transforms column stochastic $d_{A_1}\times d_{A_0}$ matrices to column stochastic $d_{B_1}\times d_{B_0}$ matrices. Indeed, let $\pi_{ai}$ satisfy $\sum_a \pi_{ai} = 1$. Then 

$$    \pi'_{bj} = \sum_{i,a} T_{jb,ia} \pi_{ai} , $$
satisfies

$$  \sum_b \pi'_{bj} = \sum_{i,a} \sum_b T_{jb,ia} \pi_{ai} =  \sum_i t_{ji} \sum_a \pi_{ai} =1 . $$
Hence, the diagonal elements of the Choi matrices of quantum channels $\Phi$  give rise to classical channels (stochastic matrices $S_{ia}$) and the diagonal elements of the Choi matrices of quantum superchannels $\Theta$   give rise to classical superchannels (matrices $T_{ia,jb}$).

{

\begin{remark}
  
We stress that a classical channel is represented by a matrix $S_{ij}$ which is column stochastic, i.e. $S_{ij} \geq 0$  together with $\sum_i S_{ij}  = 1 $ for all $j$.  Note, however, that a classical superchannel $T_{ia,jb}$ gives rise to a  matrix $t_{ji}$ which is row stochastic, i.e. $t_{ji}\geq 0$ together with $\sum_i t_{ji}=1$ for all $j$. 
  
\end{remark}

}

\section{Covariant  Supermaps}\label{sec3} 

Consider  a group $G$ and its two unitary representations $U_g$ and $V_g$ in $\HAa$ and $\HAb$, respectively. We say that a channel $\Phi : \BAa \to \BAb$ is $G$-covariant w.r.t. these representations if for all $g \in G$

\begin{equation}   \label{}
    \Phi \circ \mathcal{U}_g = \mathcal{V}_g \circ \Phi ,
\end{equation}
where
$$ \mathcal{U}_g(X) := U_g X U_g^\dagger \ , \ \ \ \ \mathcal{V}_g(X) := V_g X V_g^\dagger \ ,
$$
{define the corresponding adjoint representations. }
Equivalently, one has the following

$$ \mathcal{V}_g^\ddag \circ \Phi \circ \mathcal{U}_g = \Phi \ ,\ \ \ \ \mathcal{V}_g \circ \Phi \circ \mathcal{U}^\ddag_g = \Phi ,  $$
where $\Phi^{\ddag}$ stands for the adjoint of $\Phi$ w.r.t. the Hilbert-Schmidt inner product. 
{This covariance condition admits a natural representation-theoretic interpretation, namely, all covariant maps define the intertwiner space

\begin{equation}
\operatorname{Hom}_G\!\left(\mathcal{B}\left(\mathcal{H}^{A_0}\right),\mathcal{B}\left(\mathcal{H}^{A_1}\right)\right)
=
\left\{
\Phi \, |\, 
\Phi \circ \mathcal{U}_g
=
\mathcal{V}_g \circ \Phi,\;
\forall\, g\in G
\right\},
\end{equation}
i.e. any covariant $\Phi$ intertwines $\mathcal{U}_g$ and $\mathcal{V}_g$ representations. When $\mathcal{H}^{A_0}=\mathcal{H}^{A_1}$ and $U_g=V_g$, the intertwiner space reduces to the commutant  of the adjoint representation,
\[
\operatorname{End}_G\!\left(\mathcal{B}\left(\mathcal{H}\right)\right)
=
\left\{
\Phi \, |\, 
\Phi \circ \mathcal{U}_g
=
\mathcal{U}_g \circ \Phi,
\ \forall\, g\in G
\right\}
=
\{\mathcal{U}_g : g\in G\}'\ .
\]

}
\begin{proposition}\label{PRO-UV} {A map} $\Phi \in \LA$ is $G$-covariant iff its Choi matrix $C_\Phi$ satisfies

    \begin{equation}
  \overline{U}_g \otimes {V}_g \, C_\Phi\, \left(\overline{U}_g \otimes {V}_g\right)^\dagger = C_\Phi ,  
\end{equation}
for all $g \in G$, which means that the Choi matrix belongs to the commutant

\[ C_\Phi \in  \{ \, \overline{U}_g \otimes V_g\, \ ; \ g \in G\, \}' .  \]

\end{proposition}
We have provided a detailed proof of the above Proposition in Appendix~\ref{proof_prop3}. 
{   Therefore, the structure of covariant channels follows immediately from the decomposition of the corresponding representations into irreducible components. By Schur's lemma, an intertwiner can only connect isomorphic irreducible subrepresentations. Consequently, the classification of covariant channels reduces to determining the decomposition of the corresponding operator space into weight sectors. 
}

Now, let us generalize $G$-covariance on the level of supermaps. Let $U'_g$ and $V'_g$ be two unitary representations of $G$ in $\HBa$ and $\HBb$, respectively.

\begin{definition}
    A supermap $\Theta : \mathcal{L}^{A} \to \mathcal{L}^{B}$ is called $G$-covariant w.r.t. $\left(U,V,U',V'\right)$ iff

\begin{equation}  \label{COV1}
    \Theta[\mathcal{V}_h \circ \Phi \circ \mathcal{U}^\ddag_g ] =  \mathcal{V'}_h \circ \Theta[\Phi] \circ \mathcal{U'}^{\ddag}_g ,
\end{equation}
for all $\Phi \in \LA$ and all $\left(g,h\right) \in G \times G$. 
 
\end{definition}
Equivalently, by introducing two supermaps,

\begin{equation}
    \Omega_{\left(h,g\right)}[\Phi] := \mathcal{V}_h \circ \Phi \circ \mathcal{U}^\ddag_g \ , \quad \Omega'_{\left(h,g\right)}[\Phi] := \mathcal{V}'_h \circ \Phi \circ \mathcal{U'}^{\ddag}_g,
\end{equation}
 the formula (\ref{COV1}) leads to 

\begin{equation}
    \Theta \circ \Omega_{\left(h,g\right)} = \Omega'_{\left(h,g\right)} \circ \Theta . 
\end{equation}
For detailed discussion of covariant quantum superchannels, see Ref.~\cite{Giulio2006}.
\begin{remark} Note that for supermaps $\Theta$, or equivalently bipartite maps $\Delta_\Theta$, we have two `{input}' symmetries  $\mathcal{U}_g$ and $\mathcal{U}'_g$, and two `{output}' symmetries $\mathcal{V}_g$ and $\mathcal{V}'_g$. Hence, one may call $\Theta$ input G-covariant if

\begin{equation}  
    \Theta[\Phi \circ \mathcal{U}^\ddag_g ] =   \Theta[\Phi] \circ \mathcal{U'}^{\ddag}_g ,
\end{equation}
and output G-covariant if 
\begin{equation} 
    \Theta[\mathcal{V}_h \circ \Phi] =  \mathcal{V'}_h \circ \Theta[\Phi] .
\end{equation}
\end{remark}

\begin{proposition} \label{PRO-MN} Consider a supermap $\Theta : \LA \to \LB$, defined as

\begin{equation}
    \Theta[\Phi] = \mathcal{N}_1 \circ \Phi \circ \mathcal{N}^\ddag_0 ,
\end{equation}
with arbitrary maps $\mathcal{N}_1 :\BAb \to \BBb$ and $\mathcal{N}_0 : \BAa \to \BBa$. Then the corresponding representing map $\Delta_\Theta$ has the following form

\begin{equation}
    \Delta_\Theta = \mathbb{T}[\mathcal{N}_0] \otimes \mathcal{N}_1 ,
\end{equation}
    where 
    
\begin{equation} \label{TT}
\mathbb{T}\left[\Phi\right]:=T_{B_0}\circ\Phi\circ T_{A_0} ,     
\end{equation}    
and $T_{A_0}$ and $T_{B_0}$ denotes transpositions in $\mathcal{H}^{A_0}$ and $\mathcal{H}^{B_0}$ defined w.r.t. orthonormal bases $\{e_i\}$ and $\{f_a\}$, respectively. 
\end{proposition}

We have provided a detailed {proof} in Appendix~\ref{PROOF-MN}.
{
Observe, that $\mathcal{U}_g(X) = U_g X U^\dagger_g$ implies

\[ \mathbb{T}[\mathcal{U}_g](X)= T \circ \mathcal{U}_g \left(X^{T}\right)=\left(U_g X^{T} U_g^{\dagger}\right)^{T}=\overline{U}_g X \,U_g^{T} , \]
where $T := T_{A_0}$ (in what follows, we omit explicit reference to the underlying space whenever it is clear from the context). Hence, we  introduce the following notation:

\[  \overline{\mathcal{U}}_g(X) := \overline{U}_g X U_g^{T} .\]
}
Using Proposition \ref{PRO-MN}, one immediately finds the representing maps of $\Omega_{\left(h,g\right)}$ and $\Omega'_{\left(h,g\right)}$ as:

\begin{equation}
    \Delta_{\left(h,g\right)} = \overline{\mathcal{U}}_g \otimes \mathcal{V}_h \ , \quad 
    \Delta'_{\left(h,g\right)} = \overline{\mathcal{U}'}_g \otimes \mathcal{V}'_h \ 
\end{equation}
{The above equations can be rewritten as }
%that is

\begin{equation}
        \Delta_{\left(h,g\right)}\left(X\right) = \overline{U}_g \otimes V_h \, X \, \left(\overline{U}_g \otimes V_h\right)^\ddag \ , \quad
        \Delta'_{\left(h,g\right)}\left(X\right) = \overline{U'}_g \otimes V'_h \, X \, \left(\overline{U'}_g \otimes V_h\right)^\ddag .
\end{equation}
As a result, we prove the following:

\begin{proposition} A supermap is $G$-covariant iff its representing map satisfies

\begin{equation}
    \Delta_\Theta \circ \overline{\mathcal{U}}_g \otimes \mathcal{V}_h = \overline{\mathcal{U}'}_g \otimes \mathcal{V}'_h \circ \Delta_\Theta ,
\end{equation}
for all $(g,h) \in G \times G$.     Equivalently, for all operators $X_A \in \BAA$,
\begin{equation}
\Delta_\Theta\!\left[
  \left(\overline{U}_g \otimes V_h\right)\, X_A\, \left(\overline{U}_g \otimes V_h\right)^{\dagger}
\right]
=
\left(\overline{U}'_g \otimes V'_h\right)\,
\Delta_\Theta\left(X_A\right)\,
\left(\overline{U}_g' \otimes V'_h\right)^{\dagger}.
\end{equation}
\end{proposition}
Finally, the covariance property of $\Theta$ can be easily reformulated in terms of its Choi matrix $C_\Theta= C_\Delta$ {as follows}:

\begin{equation} \label{CC}
 \left(U_g \otimes \overline{V}_h \otimes \overline{U'}_g \otimes {V'}_h\right)\,   C_\Theta \, \left(U_g \otimes \overline{V}_h \otimes \overline{U'}_g \otimes {V'}_h\right)^\dagger = C_\Theta , 
\end{equation}
or equivalently

\begin{equation}
    W_{g,h} \otimes \overline{W}_{g,h} \, C_\Theta\, \left( W_{g,h} \otimes \overline{W}_{g,h} \right)^\dagger = C_\Theta ,
\end{equation}
with $W_{g,h} = U_g \otimes \overline{V}_h$,  {which means that the Choi matrix belongs to the commutant

\[ C_\Theta \in  \{ \, U_g \otimes \overline{V}_h \otimes \overline{U'}_g \otimes {V'}_h \, \ ; \ g,h \in G\, \}' .  \]

}
\section{Unitary Covariant Superchannels}\label{sec4} 

Let us briefly recall the structure of unitary quantum channels. %and diagonal unitary covariant  
Let $\HAb = \HAa = \mathbb{C}^d$ and $G=U(d)$ be a group of unitary $d \times d$ matrices. Let $U_g = V_g$ be a fundamental (defining) representation of $U(d)$. Then a channel $\Phi : \Md \to \Md$ is called unitary covariant if

\begin{equation} \label{U1}
    \Phi\circ \mathcal{U} = \mathcal{U} \circ \Phi ,
\end{equation}
for all unitary matrices $U \in U(d)$, that is,

\begin{equation} \label{U1X}
    \Phi({U}XU^\dagger)  = {U} \Phi\left(X\right) U^\dagger .
\end{equation}
The corresponding Choi matrix is $\left(U,\overline{U}\right)$-invariant, i.e.

\begin{equation}
    U \otimes \overline{U}\, C_\Phi\, \left(U \otimes \overline{U}\right)^\dagger = C_\Phi .
\end{equation}
Note, that the corresponding commutant

\begin{equation}
    \{ U \otimes \overline{U} \, |\, U \in U(d) \}' = {\rm span}_{\mathbb{C}} \{ \oper \otimes \oper , \mathbb{P}\} ,
\end{equation}
where $\mathbb{P} = |\psi\rangle\langle \psi|  $ and $|\psi \rangle= \sum_{i=1}^d e_i \otimes e_i$. Hence, any $(U,\overline{U})$-invariant Choi matrix reads

\[   C = \lambda \mathbb{P} + \left(1-\lambda\right) \frac 1d \oper \otimes \oper .\]
Equivalently, any unitary covariant quantum channel has the following form

\begin{equation}
    \Phi = \lambda \, {\rm id} + \left(1- \lambda\right)\, {\rm D} , \quad - \frac{1}{d^2-1} \leq \lambda \leq 1 , 
\end{equation}
where ${\rm D}(X) = \frac 1d \oper_d {\rm Tr}\, X$ is a completely depolarizing channel. 

If $V_g = \overline{U}_g$ (so called contragredient representation), then instead of (\ref{U1}), {we have}

\begin{equation}  \label{U2}
    \Phi\circ \mathcal{U} = \overline{\mathcal{U}} \circ \Phi .
\end{equation}
Equivalently,

\begin{equation}  \label{U2a}
    \Phi({U} X U^\dagger) = \overline{{U}} \Phi(X) U^T .
\end{equation}
Such maps are called conjugate-covariant. The corresponding Choi matrix is $(U,{U})$-invariant, i.e.

\begin{equation}
    U \otimes {U}\, C_\Phi\, \left(U \otimes {U}\right)^\dagger = C_\Phi ,
\end{equation}
and the corresponding commutant reads

\begin{equation}
    \{ U \otimes {U} \, |\, U \in U(d) \}' = {\rm span}_{\mathbb{C}} \{ \oper \otimes \oper , \mathbb{F}\} ,
\end{equation}
where {$\mathbb{F} = \sum_{i,j=1}^d e_{ij} \otimes e_{ji}$} denotes a flip (swap) operator. 
Hence, any conjugate covariant quantum channel has the following form

\begin{equation}
    \Phi = \mu \, {\rm T} + (1- \mu)\, {\rm D} , \quad - \frac{1}{d-1} \leq \mu \leq \frac{1}{d+1} .
\end{equation}
In particular for $\mu = - \frac{1}{d-1}$ one recovers a celebrated Holevo-Werner channel

\begin{equation}
    \Phi_{\rm HW}(\rho) = \frac{1}{d-1} \left( \oper_d {\rm Tr}\, \rho  - \rho^T \right ) .
\end{equation}
Interestingly, for $d=3$ Holevo-Werner channel provides counterexamples for
 the multiplicativity of the output $p$-norm \cite{Werner2002}, the additivity of the relative entropy of
 entanglement \cite{VollbrechtWerner2001}, and the quantum analogue of Birkhoff's theorem \cite{LandauStreater1993,Wolf2009}.
Restricting $G$ to orthogonal $d \times d$ matrices, we obtain orthogonally covariant channels

\begin{equation}
    \Phi = (1-\alpha - \beta) ~{\rm id} + \alpha\, {\rm D} + \beta\,  {\rm T} , 
\end{equation}
where {the} parameters $\alpha$ and $\beta$ satisfy

$$   \alpha \geq d|\beta|\ , \ \ \ d \left(1-\alpha - \beta\right) + \frac{\alpha}{d} + \beta \geq 0 . $$

Consider now a quantum superchannel defined in terms of its representing map 

$$   \Delta_\Theta : \Md \otimes \Md \to \Md \otimes \Md .$$
There are several ways to realize unitary invariance via formula (\ref{CC}). %Consider 

\subsection{Unitary Covariant}% $(U,V,{U},{V})$-invariance }

We call a superchannel {$\Theta$} to be $\left(U,V,U,V\right)$-invariant iff its Choi matrix {$C_{\Theta}$} displays $\left(U,V,\overline{U},\overline{V}\right)$-invariance, that is,

\begin{equation}
    U \otimes V \otimes \overline{U} \otimes \overline{V} \, C_\Theta \, \left( U \otimes V \otimes \overline{U} \otimes \overline{V}\right)^\dagger = C_\Theta ,
\end{equation}
for all $U,V \in U(d)$. {The corresponding commutant reads
\begin{eqnarray}
    &&\{  U \otimes V \otimes \overline{U} \otimes \overline{V} \, | \, U,V \in U(d)\}' = \nonumber \\
    && {\rm span}_{\mathbb{C}} \{ \oper_{A_0A_1} \otimes \oper_{B_0B_1}, \oper_{A_0B_0 } \otimes \mathbb{P}_{A_1B_1}, \mathbb{P}_{A_0B_0} \otimes \oper_{A_1 B_1}, \mathbb{P}_{A_0B_0} \otimes \mathbb{P}_{A_1B_1} \, \} \ ,
\end{eqnarray}
where $\mathbb{P}_{A_0B_0}$ stands for (unnormalized) maximally entangled projector in $\mathcal{H}^{A_0} \otimes \mathcal{H}^{B_0}$, etc. Then the representing map has the following structure:}

\begin{equation}
    \Delta_\Theta = p_0\, {\rm id} \otimes {\rm id} + p_1 \, {\rm id} \otimes {\rm D} + p_2\, {\rm D} \otimes {\rm id} + p_3\, {\rm D} \otimes {\rm D} \ . \label{eq:Rep_1}
\end{equation}

{The positivity of $C_\Delta$ is conveniently analyzed by changing the order of factors in the tensor product
\[  \mathcal{H}^{A_0} \otimes  \mathcal{H}^{A_1} \otimes  \mathcal{H}^{B_0} \otimes  \mathcal{H}^{B_1} \ \mapsto \   \mathcal{H}^{A_0} \otimes  \mathcal{H}^{B_0} \otimes  \mathcal{H}^{A_1} \otimes  \mathcal{H}^{B_1}   \] realized by the unitary operation\[  \mathcal{P} = \oper_{A_0} \otimes \mathbb{F}_{A_1B_0} \otimes \oper_{B_1} ,  \]
where \[  \mathbb{F}_{A_1B_0} : \mathcal{H}^{A_1} \otimes  \mathcal{H}^{B_0} \to \mathcal{H}^{B_0} \otimes  \mathcal{H}^{A_1} , \]
denotes the canonical swap (flip) operator. Let us defined {\em flipped} Choi matrix
\begin{equation}
    C^F_\Delta := \mathcal{P} C_\Delta \mathcal{P}^\dagger . 
\end{equation}
It is evident that $C_\Delta \geq 0$ if and only if $C^F_\Delta \geq 0$. Defining the projectors 
\begin{align}
    \Pi:=\frac{1}{d}\sum_{ij} e_{ij} \otimes e_{ij}, \quad \Pi_{\perp}:=\oper-\Pi, \label{Eq:Proj_1}
\end{align}
the flipped Choi matrix can be obtained as 
\begin{align}
   C^F_\Delta = \left(p_0 d^2 +p_1+p_2+\frac{p_3}{d^2}\right) \Pi \otimes\Pi+\left( p_1+ \frac{p_3}{d^2}\right) \Pi \otimes \Pi_{\perp}+ \left( p_2+ \frac{p_3}{d^2}\right) \Pi_{\perp} \otimes \Pi+ \frac{p_3}{d^2} \Pi_{\perp} \otimes \Pi_{\perp} . \nonumber
\end{align}
Consequently, we have the following corollary. }

\begin{corollary}
 {The representing map in Eq.~(\ref{eq:Rep_1}) is CP if and only if the coefficients satisfy the conditions }

$$p_3\geq 0\ , \ \ \ p_1 + \frac{p_3}{d^2} \geq 0\ ,\ \ \   p_2 + \frac{p_3}{d^2} \geq 0 ,  $$
and 
$$ p_0 d^2+  p_1+p_2 + \frac{p_3}{d^2} \geq 0 \ . $$   

\end{corollary}
Note that {since $\Theta$ is a quantum superchannel,} $\Delta_{\Theta}$ has to satisfy the following equation

\begin{equation}
    {\rm Tr}_{B_1} \Delta_\Theta(X) = \widetilde{\Delta}_\Theta\left({\rm Tr}_{A_1}X\right). 
\end{equation}
Using the form of $\Delta_\Theta$, we obtain the following.

\begin{equation}
    \widetilde{\Delta}_\Theta = (p_0 + p_1)\, {\rm id} + (p_2 + p_3)\, {\rm D} ,
\end{equation}
which is a unitary covariant unital quantum channel. 
Taking into account that ${\rm D}^\ddag = {\rm D}$ and $\mathbb{T}[{\rm D}]={\rm D}$ this yields

\begin{equation}
    \Theta[\Phi] = p_0\, \Phi + p_1 \, {\rm D} \circ \Phi + p_2 \, \Phi \circ {\rm D} + p_3\, {\rm D} \circ \Phi \circ {\rm D} . 
\end{equation}

\subsection{Conjugate Unitary Covariant}

We call a superchannel $(U,V,\overline{U},\overline{V})$-invariant iff its Choi matrix displays $\left(U,V,U,V\right)$-invariance, that is,

\begin{equation}
    U \otimes V \otimes {U} \otimes {V} \, C_\Theta \, \left( U \otimes V \otimes {U} \otimes {V}\right)^\dagger = C_\Theta .
\end{equation}
{The representing map of $\Theta$ has the following structure:}

\begin{equation}
    \Delta_\Theta = p_0\, {\rm T} \otimes {\rm T} + p_1 \, {\rm T} \otimes {\rm D} + p_2\, {\rm D} \otimes {\rm T} + p_3\, {\rm D} \otimes {\rm D} \ . \label{eq:Rep_2}
\end{equation}
{Observing that $C_T=\mathbb{F}$, and defining the projectors 
\begin{align}
    \Pi_+:=\frac{1}{2}(\oper \otimes \oper +\mathbb{F}), \quad  \Pi_-:=\frac{1}{2}(\oper\otimes \oper -\mathbb{F}), \label{Eq:Proj_2}
\end{align}
satisfying $\Pi_+ + \Pi_-= \oper \otimes \oper$, 
we arrive at the following form of the flipped Choi matrix $C^F_\Delta$:
\begin{align}
    C^F_\Delta &=\left( p_0+\frac{p_1}{d}+ \frac{p_2}{d}+\frac{p_3}{d^2}\right) \Pi_+ \otimes \Pi_+ + \left( -p_0+\frac{p_1}{d}- \frac{p_2}{d}+\frac{p_3}{d^2}\right) \Pi_+ \otimes\Pi_- \nonumber\\
    &\quad+ \left( -p_0-\frac{p_1}{d}+ \frac{p_2}{d}+\frac{p_3}{d^2}\right) \Pi_- \otimes \Pi_+ \left( p_0-\frac{p_1}{d}- \frac{p_2}{d}+\frac{p_3}{d^2}\right) \Pi_- \otimes \Pi_- . 
\end{align}
The above decomposition leads to the following necessary and sufficient condition for the complete positivity of 
$\Delta_{\Theta}$:
\begin{corollary}
The representing map in Eq.~(\ref{eq:Rep_2}) is CP if and only if 
%$\Delta_{\Theta}$:
%Observe that $\Delta_{\Theta}$ is CP if and only if the coefficients $p_\alpha$ satisfy the following conditions 
\begin{equation}
   \frac{p_3}{d^2} \pm p_0 \geq \frac{1}{d}\,  \vert p_1 \pm p_2 \vert .
\end{equation}   
\end{corollary}
}
Note that

\begin{equation}
    {\rm Tr}_{B_1} \Delta_\Theta(X) = \widetilde{\Delta}_\Theta({\rm Tr}_{A_1}X) ,
\end{equation}
where

\begin{equation}
    \widetilde{\Delta}_\Theta = (p_0 + p_1)\, {\rm T} + (p_2 + p_3)\, {\rm D} ,
\end{equation}
is a conjugate unitary covariant unital quantum channel.  
Taking into account that ${\rm T}^\ddag = {\rm T}$ and $\mathbb{T}[{\rm T}]={\rm T}$ we obtain

\begin{equation}
    \Theta[\Phi] = p_0\, \mathbb{T}[\Phi] + p_1 \, {\rm D} \circ \Phi \circ {\rm T} + p_2 \,  {\rm T} \circ \Phi \circ {\rm D} + p_3\,  {\rm D} \circ \Phi \circ {\rm D} . 
\end{equation}
In particular, one defines the Holevo-Werner superchannel

\begin{equation}
    \Theta_{\rm HW}[\Phi] = \frac{1}{d^2-1} \Big( d^2 \, {\rm D} \circ \Phi \circ {\rm D} - \mathbb{T}[\Phi] \Big) . 
\end{equation}

\subsection{Input Unitary and Output Conjugate Unitary Covariant}

Such covariant maps are defined by the following condition
\begin{equation}
    U \otimes V \otimes \overline{U} \otimes {V} \, C_\Theta \, ( U \otimes V \otimes \overline{U} \otimes {V})^\dagger = C_\Theta ,
\end{equation}
for all $U,V \in U(d)$. One derives the following structure of the representing map
\begin{equation}
    \Delta_\Theta = p_0\, {\rm id} \otimes {\rm T} + p_1 \, {\rm id} \otimes {\rm D} + p_2\, {\rm D} \otimes {\rm T} + p_3\, {\rm D} \otimes {\rm D} \,. \label{eq:rep_3}
\end{equation}
{In terms of the projectors defined in Eq.~(\ref{Eq:Proj_1}) and Eq.~(\ref{Eq:Proj_2}), the flipped Choi matrix can be written as
\begin{align}
    C^F_\Delta &=\left( p_0 d+p_1+ \frac{p_2}{d}+\frac{p_3}{d^2}\right) \Pi \otimes \Pi_+ +  \left( -p_0 d+p_1- \frac{p_2}{d}+\frac{p_3}{d^2}\right) \Pi \otimes\Pi_- \nonumber\\
    & \quad +  \left(\frac{p_2}{d}+\frac{p_3}{d^2}\right) \Pi_{\perp} \otimes \Pi_+ +  \left(-\frac{p_2}{d}+\frac{p_3}{d^2}\right) \Pi_{\perp} \otimes \Pi_-.
\end{align}
}
{The above decomposition leads to the following necessary and sufficient condition for the complete positivity of 
$\Delta_{\Theta}$:}

{
\begin{corollary}
The representing map $\Delta_{\Theta}$ in Eq.~(\ref{eq:rep_3}) is CP if and only if the coefficients $p_\alpha$ satisfy the following conditions
\begin{equation}
    p_3 \geq d \vert p_2 \vert, ~~~ d p_1 + \frac{p_3}{d} \geq  \vert d^2 p_0  + p_2 \vert.
\end{equation}
\end{corollary} }

\section{Diagonal Unitary Covariant Superchannels}\label{sec5}

Consider now the group of diagonal unitary matrices in $\mathcal{M}_d$, i.e. $U = \sum_k e^{i \phi_k} e_{kk} $ with $\phi_k \in [0,2\pi)$. {Clearly, it is topologically equivalent to  a $d$-dimensional torus

\[  T_d = \{ (\phi_1,\ldots,\phi_d)\ |\  \phi_k \in [0,2\pi) \}  . \]
}
One proves \cite{Singh_2021} the following (cf. also \cite{DC_AK_2006, Chruscinski2022Dynamical ,DC_Topical,DC_AK_OSID})

\begin{proposition}   \label{PRO-6}
Linear maps covariant w.r.t. the group of diagonal unitary matrices have the following form 

\begin{equation}   \label{Phi-AB}
    \Phi(X) = \sum_{i,j=1}^d A_{ij}\, e_{ij} X e_{ij}^\dagger + \sum_{i \neq j} {B}_{ij} \,e_{ii} X e_{jj} ,
\end{equation}
where $A$ and $B$ are arbitrary $d \times d$ matrices with ${\rm diag}\, B=0$. 
\end{proposition}
{
\begin{proof} The proof  easily follows from the decomposition of the operator space $\Md$ into weight spaces for the  action of $T_d$. One has

\[  \mathcal{U}(e_{k\ell}) = e^{i(\phi_k - \phi_\ell)} e_{k\ell} , \]
and hence for  $e_{k\ell}$ is a weight vector with the corresponding weight 
$ \chi_{k\ell} = e^{i(\phi_k - \phi_\ell)}$. 
One derives, therefore, the following decomposition of the operator space

\[  \Md = V_0 \oplus \bigoplus_{k\neq \ell} V_{k\ell} ,\]
where

\[ V_{k\ell}= \mathbb{C}\, e_{k\ell} \ , \ \ (k \neq \ell) ; \ \ \ \ \ \ V_0 = {\rm span}_\mathbb{C} \{e_{11},\ldots,e_{dd} \} . \]
The weight 1 appears with multiplicity $d$. Now, due to the Schur lemma, the intertwinner $\Phi$ can connect only the spaces with the same weights, one has $\Phi(V_0) \subset V_0$ and $\Phi(V_{k\ell}) \subset V_{k\ell}$ and hence one finds immediately

\[  \Phi(e_{k\ell}) = B_{k\ell}\, e_{k\ell} \ ,\]
for $k\neq \ell$, and for the space with weight 1 

\[   \Phi(e_{kk}) = \sum_{\ell} A_{\ell k}\, e_{\ell\ell} , \]
which is equivalent to (\ref{Phi-AB}). 
\end{proof}}

The map $\Phi$ defined in (\ref{Phi-AB}) is
\begin{enumerate}
    \item Hermiticity preserving if $A_{ij} \in \mathbb{R}$ and $B_{ij} = \overline{B}_{ji}$ ,
    \item completely positive if $A_{ij} \geq 0$ and the following matrix

$$   \mathbb{B}_{ij} = \left\{ \begin{array}{ll} {B}_{ij} \ & \ i \neq j \\
A_{ii} & \ i=j \end{array}  \right. $$
is positive definite,

\item trace preserving iff $\sum_{i}A_{ij}=1$. 
\end{enumerate}
Hence, if $\Phi$ represents a quantum channel then the matrix $A_{ij}$ represents a classical channel.

Similarly, a  conjugate covariant (w.r.t. diagonal unitaries) linear map has the following form \cite{Singh_2021}

\begin{equation}
    \Phi(X) = \sum_{i,j=1}^d A_{ij}\, e_{ij} X e_{ij}^\dagger + \sum_{i \neq j} C_{ij}\, e_{ii} X^T e_{jj} .
\end{equation}
Any such map defines a quantum channel iff $A_{ij}$ is a column stochastic matrix, and $C_{ij} = \overline{C_{ji}}$ together with  the following positivity condition

\begin{equation}
    |C_{ij}|^2 \leq A_{ij} A_{ji} \ , \ \ i \neq j . 
\end{equation}
Finally, channels covariant w.r.t. diagonal orthogonal matrices have the following structure \cite{Singh_2021}

\begin{equation}
    \Phi(X) = \sum_{i,j=1}^d A_{ij}\, e_{ij} X e_{ij}^\dagger + \sum_{i \neq j} B_{ij}\, e_{ii} X e_{jj} + \sum_{i \neq j} C_{ij}\, e_{ii} X^T e_{jj} .
\end{equation}

Consider now a quantum superchannel  $\Theta : \LA \to \LB$ together with its representing map $\Delta : \Md \otimes \Md \to \Md \otimes \Md$. $G$ is a group of diagonal $d \times d$ unitary matrices. To distinguish between elements in $\Md \otimes \Md$, we enumerate  matrix units in the first factor by $e_{ij}$ and in the second factor by $e_{ab}$.

\begin{theorem}\label{thm1}
A supermap which is {\rm DU}-covariant has the following form
\begin{equation}
    \Delta = \Delta_1 + \Delta_2 + \Delta_3 + \Delta_4 , \label{DUC_SUPMAP} 
\end{equation}
where
\begin{eqnarray}\label{Delta-1}
    \Delta_1(X) &=& \sum_{i,j} \sum_{a,b} A_{ia,jb}\ e_{ij} \otimes e_{ab}\, X\, e_{ji} \otimes e_{ba} \ , \nonumber\\
    \Delta_2(X) &=& \sum_{i,j} \sum_{a\neq b} B_{ia,jb}\ e_{ij} \otimes e_{aa} \, X\, e_{ji} \otimes e_{bb} \ , \\
    \Delta_3(X) &=& \sum_{i\neq j} \sum_{a,b} C_{ia,jb}\ e_{ii} \otimes e_{ab} \, X\, e_{jj} \otimes e_{ba} \ , \nonumber\\
    \Delta_4(X) &=& \sum_{i\neq j} \sum_{a\neq b} D_{ia,jb}\  e_{ii} \otimes e_{aa} \, X\, e_{jj} \otimes e_{bb} \nonumber\ .
\end{eqnarray}  
\end{theorem}
{\begin{proof}
    
 Let $U = \sum_j u_j \, e_{jj}$ and $V = \sum_a v_a \, e_{aa}$ be two diagonal unitaries from $U(d)$, where

\[   u_j = e^{i\phi_j} \ , \ \ \ \ v_a = e^{i \psi_a} , \]
and $(\phi_1,\ldots,\phi_d), (\psi_1,\ldots,\psi_d) \in T_d$. 
It follows that

\begin{equation}
    [\mathcal{U} \otimes \mathcal{V}](e_{ij} \otimes e_{ab}) = \chi_{ij;ab} \, e_{ij} \otimes e_{ab} ,
\end{equation}
with weight

\[   \chi_{ij;ab} = {u_i} \overline{u_j} v_a \overline{v}_b . \]
%e^{i[(\phi_i - \phi_j) + (\psi_a - \psi_b)]} . \]
 This way  four weight sectors appear:

$$ V_{00} = \operatorname{span}_{\mathbb{C}} \{ e_{ii} \otimes e_{aa} : 1 \leq i, a \leq d \}, $$
with weight  $\chi_{0;0} = 1$. Next,

\[
\mathcal V_{0;ab}
=
\operatorname{span}_\mathbb{C}
\{\,e_{ii}\otimes e_{ab}:1\le i\le d\,\},
\qquad a\neq b,
\]
with weight $\chi_{0;ab} = v_a\overline{v_b}$. Similarly,

\[
V_{ij;0}
=
\operatorname{span}_\mathbb{C}
\{\,e_{ij}\otimes e_{aa}:1\le a\le d\,\},
\qquad i\neq j,
\]
with weight $\chi_{ij;0} =
u_i\overline{u_j}$. Finally,

\[
V_{ij;ab}
=
\mathbb C\,(e_{ij}\otimes e_{ab}),
\qquad
i\neq j,\quad a\neq b,
\]
with weight   $\chi_{ij;ab}
=
u_i\overline{u_j}\,
v_a\overline{v_b}$. The corresponding multiplicities read

\[
\dim  V_{00}=d^2, \ \ 
\dim V_{0;ab}= \dim V_{ij;0}=d,
\ \ 
\dim V_{ij;ab}=1.
\]
One derives the following decomposition of the  operator space $\Md \otimes \Md$

\[
\Md\otimes \Md
=
V_{00}
\oplus
\bigoplus_{a\neq b}
V_{0;ab}
\oplus
\bigoplus_{i\neq j}
V_{ij;0}
\oplus
\bigoplus_{\substack{i\neq j\\ a\neq b}}
V_{ij;ab}.
\]
Now, the intertwiner $\Delta$ can only  connect basis elements with the same weight. Let us introduce four operators

\begin{eqnarray*}
\Delta_1:
e_{jj}\otimes e_{bb}
& \longmapsto&
\sum_{i,a}
A_{ia,jb}\,
e_{ii}\otimes e_{aa},
\\
\Delta_2:
e_{jj}\otimes e_{ab}
& \longmapsto&
\sum_i
B_{ia,jb}\,
e_{ii}\otimes e_{ab},
\qquad a\neq b,
\\
\Delta_3:
e_{ij}\otimes e_{bb}
& \longmapsto&
\sum_a
C_{ia,jb}\,
e_{ij}\otimes e_{aa},
\qquad i\neq j,
\\
\Delta_4:
e_{ij}\otimes e_{ab}
& \longmapsto &
D_{ia,jb}\,
e_{ij}\otimes e_{ab},
\qquad i\neq j,\quad a\neq b.
\end{eqnarray*}
It is, therefore, clear that $\Delta_1 : V_{0;0} \to V_{0;0}$ and

\begin{eqnarray*}
    %&& \Delta_1 : V_{0;0} \to V_{0;0} , \\
    \Delta_2 : \bigoplus_{a\neq b}
V_{0;ab} & \longmapsto& \bigoplus_{a\neq b}
V_{0;ab} \ , \\
\Delta_3 : \bigoplus_{i\neq j}
V_{ij;0} & \longmapsto& \bigoplus_{i\neq j}
V_{ij;0} \ , \\
 \Delta_4 : \bigoplus_{\substack{i\neq j\\ a\neq b}}
V_{ij;ab} & \longmapsto& \bigoplus_{\substack{i\neq j\\ a\neq b}}
V_{ij;ab}.
\end{eqnarray*}
This way we proved that any intertwiner $\Delta$ can be uniquely decomposed as $\Delta= \sum_\alpha \Delta_\alpha$. 
\end{proof}

The covariance condition identifies the representing map $\Delta_\Theta$ as 
an intertwiner of the induced representation of the group of diagonal unitary operators acting on $\Md \otimes \Md$. Consequently, the characterization of covariant superchannels reduces to 
determining the commutant of this representation subject to the causality 
constraints defining one-slot quantum combs.

}

Introducing the following maps

\begin{equation}
    \mathcal{E}_{ij}(X) := e_{ij} X e_{ji} \ ,\quad \mathcal{E}^i_{\ j}(X) := e_{ii} X e_{jj} , 
\end{equation}
one finds the following representation for the diagonal unitary covariant supermap

$$   \Theta = \Theta_1 + \Theta_2 + \Theta_3 + \Theta_4 ,$$
where

\begin{eqnarray}  \label{Theta-1}
    \Theta_1[\Phi] &=& \sum_{i,j} \sum_{a,b} A_{ia,jb}\, \mathcal{E}_{ab} \circ \Phi \circ \mathcal{E}_{ij} \nonumber\ , \\
     \Theta_2[\Phi] &=& \sum_{i,j} \sum_{a\neq b} B_{ia,jb}\, \mathcal{E}^{a}_{\ b} \circ \Phi \circ \mathcal{E}_{ij} \ , \\
     \Theta_3[\Phi] &=& \sum_{i\neq j} \sum_{a, b} C_{ia,jb}\, \mathcal{E}_{ab} \circ \Phi \circ \mathcal{E}^{i}_{\ j} \ , \nonumber\\
      \Theta_4[\Phi] &=& \sum_{i\neq j} \sum_{a\neq  b} D_{ia,jb}\, \mathcal{E}^{a}_{\ b} \circ \Phi \circ \mathcal{E}^{i}_{\ j}\nonumber \ .
\end{eqnarray}

\begin{proposition}\label{propp6}
    A {\rm DU}-covariant supermap $\Theta$ is TP preserving iff there exist $\alpha_{ij}$ and $\gamma_{ij}$ such that for any $b=1,\ldots,d$ 
    \begin{eqnarray}   \label{TP1}
    \sum_a A_{ia,jb} = \alpha_{ij} \ , \ \ 
  %  B_{ia,ja} = \beta_{ij} \ , \ \ 
    \sum_a C_{ia,jb} = \gamma_{ij} \ , \ \
 %   D_{ia,ja} = \kappa_{ij} \ ,
\end{eqnarray}
together with the following normalization condition
\begin{equation}
    \sum_j \alpha_{ij} =1 \ ,% \beta_{ij} ) + \gamma_{ii} + \kappa_{ii} = 1 , 
\end{equation}
for all $i=1,\ldots,d$.  
\end{proposition}
{\begin{proof}
    
  A supermap $\Theta$ is TP preserving if and only if the corresponding representing map $\Delta$ satisfies the following condition:  there exists a map $\widetilde{\Delta} : \mathcal{B}(\mathcal{H}^{A_0}) \to \mathcal{B}(\mathcal{H}^{B_0}) $ such that

\begin{equation}  \label{TPTP}
 {\rm Tr}_{B_1} \circ \Delta = \widetilde{\Delta} \circ {\rm Tr}_{A_1} .     
\end{equation}
 Consider the following decomposition of a bipartite operator $X \in \mathcal{B}(\mathcal{H}^{A_0A_1})$ as $X = \sum_{cd} X_{cd}\otimes e_{cd} $, with $X_{cd} \in \mathcal{B}(\mathcal{H}^{A_0})$.  One has

\[   {\rm Tr}_{B_1} \circ \Delta (X) = \sum_\alpha  {\rm Tr}_{B_1} \circ \Delta_\alpha (X) =  {\rm Tr}_{B_1} \circ \Delta_1 (X) +  {\rm Tr}_{B_1} \circ \Delta_3 (X) , \]
since

\[   {\rm Tr}_{B_1} \circ \Delta_2(X) =  {\rm Tr}_{B_1} \circ \Delta_4(X) = 0 ,\]
for all $X \in \mathcal{B}(\mathcal{H}^{A_0A_1})$. Hence
\begin{equation}
  {\rm Tr}_{B_1} \circ \Delta (X) =   \sum_b
\sum_{i,j}
\left( \sum_a A_{ia,jb} \right)
e_{ij} X_{bb} e_{ji}
+
\sum_b
\sum_{i\neq j}
\left( \sum_a C_{ia,jb} \right)
e_{ii} X_{bb} e_{jj}.
\end{equation}
Now, since ${\rm Tr}_{A_1}X = \sum_b X_{bb}$, the condition (\ref{TPTP}) is satisfied if and only if there exist matrices $\alpha=(\alpha_{ij})$
and $\gamma=(\gamma_{ij})$, such that
$ \sum_a A_{ia,jb} = \alpha_{ij},~
\sum_a C_{ia,jb} = \gamma_{ij},$ independent of $b$. This way the map $\widetilde{\Delta}$ is defined as follows

\begin{equation}
    \widetilde{\Delta}(A) = \sum_{i,j=1}^d \alpha_{ij}\, e_{ij} A e_{ji} + \sum_{i\neq j} \gamma_{ij}\, e_{ii} A e_{jj} ,
\end{equation}
for $A \in \mathcal{B}(\mathcal{H}^{A_0})$. Finally,   the unitality condition of $\widetilde{\Delta}$ implies 

\[    \widetilde{\Delta}(\oper)  = \sum_{i,j=1}^d \alpha_{ij}\, e_{ii} + \sum_{i\neq j} \gamma_{ij}\, e_{ii} e_{jj} = \sum_{i,j=1}^d \alpha_{ij}\, e_{ii} = \oper , \]
if and only if $\sum_{j} \alpha_{ij} = 1$. 
This completes our proof. 

\end{proof}
\begin{remark} Note, that the map $\widetilde{\Delta}$ is DU-covariant, i.e. it belongs to the class (\ref{Phi-AB}).
    
\end{remark}

}

Finally, let us analyze when $\Theta$ is completely CP-preserving, or, equivalently, when the corresponding representing map $\Delta$ is completely positive. The Choi matrix of $\Theta$ reads

\begin{eqnarray*}
C_\Theta &=&  \sum_{i,j} \sum_{a,b} A_{ia,jb}\ e_{jj} \otimes e_{bb} \otimes e_{ii} \otimes e_{aa} +
 \sum_{i,j} \sum_{a\neq b} B_{ia,jb}\ e_{jj} \otimes e_{ab} \otimes e_{ii} \otimes e_{ab}  \\
 &+&   \sum_{i\neq j} \sum_{a, b} C_{ia,jb}\ e_{ij} \otimes e_{bb} \otimes e_{ij} \otimes e_{aa} +
 \sum_{i\neq j} \sum_{a\neq b} D_{ia,jb} \ e_{ij} \otimes e_{ab} \otimes e_{ij} \otimes e_{ab} .
\end{eqnarray*}
In the simplest scenario corresponding to $d=2$, we obtain the following $d^4 \times d^4$ matrix:

\begin{equation}   \label{C-d=2}
C_{\Delta} =    \left(
\begin{array}{cccc|cccc|cccc|cccc}
 A & . & . & . & . & B & . & . & . & . & C & . & . & . & . & D \\
 . & A & . & . & . & . & . & . & . & . & . & C & . & . & . & . \\
 . & . & A & . & . & . & . & B & . & . & . & . & . & . & . & . \\
 . & . & . & A & . & . & . & . & . & . & . & . & . & . & . & . \\ \hline
 . & . & . & . & A & . & . & . & . & . & . & . & . & . & C & . \\
 B & . & . & . & . & A & . & . & . & . & D & . & . & . & . & C \\
 . & . & . & . & . & . & A & . & . & . & . & . & . & . & . & . \\
 . & . & B & . & . & . & . & A & . & . & . & . & . & . & . & . \\   \hline
 . & . & . & . & . & . & . & . & A & . & . & . & . & B & . & . \\
 . & . & . & . & . & . & . & . & . & A & . & . & . & . & . & . \\
 C & . & . & . & . & D & . & . & . & . & A & . & . & . & . & B \\
 . & C & . & . & . & . & . & . & . & . & . & A & . & . & . & . \\   \hline
 . & . & . & . & . & . & . & . & . & . & . & . & A & . & . & . \\
 . & . & . & . & . & . & . & . & B & . & . & . & . & A & . & . \\
 . & . & . & . & C & . & . & . & . & . & . & . & . & . & A & . \\
 D & . & . & . & . & C & . & . & . & . & B & . & . & . & . & A \\
\end{array} \right) ,
\end{equation}
where to make the structure of the matrix more transparent, we have replaced all zeros by dots and skipped the indices of matrix elements of $\{A,B,C,D\}$. A supermap $\Delta_\Theta$ is Hermiticity preserving iff  $C_\Delta^\dagger = C_\Delta$, that is,  $A_{ia,jb} \in \mathbb{R}$ together with

\begin{equation}   \label{BCD-Her}
     B_{ia,jb} = \overline{B_{ib,ja}} \ , \quad C_{ia,jb} = \overline{C_{ja,ib}}\ , \quad D_{ia,jb} = \overline{D_{jb,ia}}.
\end{equation}

{

\begin{proposition} A DU-covariant supermap $\Delta$ is parametrized by $N_{\rm DU}=d^2(2d-1)^2$ complex parameters. If $\Delta$ is also Hermiticity-preserving, then it is parametrized by the same number of real parameters. If additionally the TP preservation condition (\ref{TP1}) is satisfied, then $N_{\rm DU}^{\rm TP} = 2d(d-1)(2d^2-d+1)$.
\end{proposition}
\begin{proof}
  
The number of independent complex parameters encoded into $(A,B,C,D)$ reads

\begin{eqnarray*}
A_{ia,jb} &:& \quad d^4, \\
B_{ia,jb}  &:&  \quad d^2\,d(d-1)=d^3(d-1), \\
C_{ia,jb}  &:&  \quad d(d-1)\,d^2=d^3(d-1), \\
D_{ia,jb}  &:&  \quad d(d-1)\,d(d-1)=d^2(d-1)^2.
\end{eqnarray*}
Hence, the total number of independent complex parameters equals
\[
N_{\rm DU}
=
d^4
+
2d^3(d-1)
+
d^2(d-1)^2
=
d^2(2d-1)^2.
\]
A Similar count of parameters applies to Hermiticity preserving maps, i.e. when $A$ is a real matrix and $(B,C,D)$ satisfy (\ref{BCD-Her}). In this case, one has $N_{\rm DU}$ real parameters. Adding the TP conditions (\ref{TP1}), we have additionally $d^2(d-1)+d$ for $A_{ia,jb}$ and $(d-1)(d^2-d)$ constraints for $C_{is,jb}$. Hence

\[ N_{\rm DU}^{\rm TP} = 2d(d-1)(2d^2-d+1)  \]
which completes the proof. 
\end{proof}

\begin{remark}
Note that for $d=2$, we have $N_{DU}=36$, which perfectly agrees with the total number of nonvanishing matrix elements in (\ref{C-d=2}).  Interestingly, the number of independent complex parameters of any DU-covariant map $\Md\to \Md$ equals to $n_{\rm DU}= d^2 + d(d-1) = d(2d-1)$ and hence $N_{\rm DU}=n_{\rm DU}^2$. 
\end{remark}
}

{To analyze the positivity of $C_\Delta$ it is convenient to change the order of the factors in the tensor product

\[  \mathcal{H}^{A_0} \otimes  \mathcal{H}^{A_1} \otimes  \mathcal{H}^{B_0} \otimes  \mathcal{H}^{B_1} \ \mapsto \   \mathcal{H}^{A_0} \otimes  \mathcal{H}^{B_0} \otimes  \mathcal{H}^{A_1} \otimes  \mathcal{H}^{B_1}   \]
which is realized by the unitary operation

\[  \mathcal{P} = \oper_{A_0} \otimes \mathbb{F}_{A_1B_0} \otimes \oper_{B_1} ,  \]
where

\[  \mathbb{F}_{A_1B_0} : \mathcal{H}^{A_1} \otimes  \mathcal{H}^{B_0} \to \mathcal{H}^{B_0} \otimes  \mathcal{H}^{A_1} , \]
denotes the canonical swap (flip) operator. Define the flipped Choi matrix 

\begin{equation}
    C^F_\Delta = \sum_{i,j=1}^d e_{ii} \otimes e_{jj} \otimes M_{ij} + \sum_{i \neq j} e_{ij} \otimes e_{ij} \otimes N_{ij} ,
\end{equation}
where $M_{ij}, N_{ij} \in \mathcal{B}(\mathcal{H}^{A_1B_1}) $ are given by

\begin{align}
M_{ij}
&=
\sum_{a,b=1}^{d}
A_{ja,\,ib}\,
e_{bb}\otimes e_{aa}
+
\sum_{\substack{ a\neq b=1}}^{d}
B_{ja,\,ib}\,
e_{ab}\otimes e_{ab},
\\[1ex]
N_{ij}
&=
\sum_{a,b=1}^{d}
C_{ia,\,jb}\,
e_{bb}\otimes e_{aa}
+
\sum_{\substack{a\neq b=1}}^{d}
D_{ia,\,jb}\,
e_{ab}\otimes e_{ab}.
\end{align}

}

\begin{proposition}\label{propp7} A {\rm DU}-covariant map $\Delta$ is completely positive iff 
   \begin{itemize}
    \item $M_{ij} \geq 0 $ for $i \neq j$, and
    \item the following operator

$$ K=  \sum_{i=1}^d e_{ii} \otimes e_{ii}\otimes M_{ii} + \sum_{i\neq j} e_{ij} \otimes e_{ij} \otimes N_{ij}  , $$
satisfies $K \geq 0$.
\end{itemize} 
\end{proposition}
{
\begin{proof} The operator
\[
C_\Delta^F
=
\sum_{i,j=1}^d
e_{ii}\otimes e_{jj}\otimes M_{ij}
+
\sum_{i\neq j}
e_{ij}\otimes e_{ij}\otimes N_{ij}
\]
admits the following invariant subspace decomposition:  define
\[
\mathcal W_{\rm off}
=
\sum_{i \neq j} \mathbb C (e_i \otimes e_j) \otimes \mathcal{H}^{A_1B_1} ,
\]
and

\[
\mathcal W_{\mathrm{diag}}
=
\operatorname{span}_{\mathbb C} \{e_{1}\otimes e_1,\ldots,e_{d}\otimes e_d\}
\otimes
\mathcal{H}^{A_1B_1} ,
\]
such that

\[  \mathcal{H}^{A_0} \otimes  \mathcal{H}^{B_0} \otimes  \mathcal{H}^{A_1} \otimes  \mathcal{H}^{B_1} = \mathcal{W}_{\rm off} \oplus \mathcal{W}_{\rm diag} . \]
Consequently,
\[
C_\Delta^F
\cong
K\,
\oplus\,
\bigoplus_{i\neq j}
M_{ij},
\]
and therefore $
C_\Delta^F\ge0$ iff $K\geq 0$ {and} $M_{ij}\ge0$ {for all } $i\neq j$.
\end{proof}

For example, in the 2-qubit case

\begin{equation*}
     {C}^F_\Delta = \begin{pmatrix} M_{11} & 0 & 0 & N_{12} \\ 0 & M_{21} & 0 & 0 \\ 0 & 0 & M_{12} & 0 \\ N_{21} & 0 & 0 & M_{22} \end{pmatrix} = K \oplus \Big(M_{12} \oplus M_{21}\Big) , 
   %   ,
\end{equation*}
where

\[   K = \begin{pmatrix} M_{11} &  N_{12} \\  N_{21} &  M_{22} \end{pmatrix} .\]

}

\begin{corollary} If $\Theta$ is a diagonal unitary covariant quantum superchannel then the corresponding matrix $A_{ia,jb}$ defines a classical superchannel. 
\end{corollary}

Let us recall that a linear space of unitary covariant maps is closed under composition \cite{Singh_2021}.

\begin{proposition} \label{PRO-Comp} Consider two diagonal unitary covariant maps

\begin{equation}
    \Phi(X) = \sum_{i,j} A_{ij} e_{ij} X e_{ji} + \sum_{i\neq j} {B}_{ij} e_{ii} X e_{jj} , \ \ 
    \Phi'(X) = \sum_{i,j} A'_{ij} e_{ij} X e_{ji} + \sum_{i\neq j} {B}'_{ij} e_{ii} X e_{jj}. 
\end{equation}
Then $\widetilde{\Phi} = \Phi \circ \Phi'$ is diagonal unitary covariant, with matrices $\widetilde{A}$ and $\widetilde{B}$
defined by:

\begin{equation}
    \widetilde{A} = A A' \ , \ \ \ \  \widetilde{B} = B \odot B' ,
\end{equation}
where $B \odot B'$ denotes the Hadamard product. 

\end{proposition}

A similar result holds for covariant supermaps. 

\begin{proposition} Consider two diagonal unitary covariant supermaps $\Theta$ and $\Theta'$, parameterized
by matrices $\{A,B,C,D\}$ and  $\{A',B',C',D'\}$, respectively. Then $\widetilde{\Theta} = \Theta \circ \Theta'$ is a diagonal unitary covariant supermap parameterized by the matrices $\{\widetilde{A},\widetilde{B},\widetilde{C},\widetilde{D}\}$, where  

\begin{equation}   \label{DD'}
    \widetilde{A} = A A' \ , \ \ \  \widetilde{D} = D \odot D' ,
\end{equation}
together with
\begin{align}
%\boxed{\;
%\widetilde{A}_{ia,jb}
%=\sum_{k}\sum_{c}\! A_{ia,kc}\,A'_{kc,jb}
%\;+\;\sum_{k}\sum_{c\neq b}\! A_{ia,kc}\,B'_{kc,jb}
%\;}\label{eq:Aupdate}\\[6pt]
%\boxed{\;
\widetilde{B}_{ia,jb}
= \sum_k B_{ia,kb}\,B'_{ka,jb} \ ,
%\ \ (a\neq b) \ ;
%\;}\label{eq:Bupdate}\\[6pt]
%\boxed{\;
\quad \widetilde{C}_{ia,jd}
=\sum_{b}\! C_{ia,jb}\,C'_{ib,jd}\ .
%\;+\;\sum_{b\neq d}\! C_{ia,jb}\,D'_{ib,jd}
%\ \ (i\neq j) .
%\;}\label{eq:Cupdate}\\[6pt]
%\boxed{\;
%\widetilde{D}_{ia,jb}
%= D_{ia,jb}D'_{ia,jb}
%\qquad(i\neq j,\ a\neq b)
%\;}\label{eq:Dupdate}
\end{align}
%i.e.\ the $D$-block is updated by the {Hadamard product}.
\end{proposition}
The proof is based on the simple observation that $ \Delta_{\alpha} \circ \Delta'_{\beta} =0$
for $\alpha \neq \beta$. Note that, representing $B$ and $C$ matrices via

$$   B = \sum_{i,j} e_{ij} \otimes b_{ij} \ , \quad C= \sum_{a,b} c_{ab} \otimes e_{ab} ,$$
with $b_{ij},c_{ab} \in \Md$, one obtains

\begin{equation}
    \tilde{b}_{ij} = \sum_k b_{ik} \odot b'_{kj} \ , \quad   \tilde{c}_{ad} = \sum_b c_{ab} \odot c'_{bd} \ .
\end{equation}

%%%%%%%%%%%%%%%%%%%%%%%%%%%%%%%%%%%%%%%%%%%%%%%%%%%%%%%%%%

\section{ Diagonal orthogonally covariant  quantum superchannels} \label{APP-ORT}

In this section, we study supermaps which are covariant with respect to diagonal orthogonal group. Let $DO$ denote the set of diagonal orthogonal matrices in $\Md$. A linear map $\Phi: \Md \to \Md$ is diagonal orthogonal covariant iff it displays the following form ~\cite{Singh_2021}
\begin{equation}  
   \Phi(X) = \sum_{i, j} A_{ij} e_{ij} X e_{ji} + \sum_{i \neq j} B_{ij} e_{ii} X e_{jj} + \sum_{i \neq j} C_{ij} e_{ii} X^T e_{jj} . \label{equ:diagonal_orthogonal_map}
\end{equation} 
Complete positivity conditions can be characterize in terms  matrices $A,B,C$ as follows: 
  \begin{itemize}
    \item $\Phi$ is Hermiticity preserving if and only if $ A_{ij} \in \mathbb{R}$,  $ B_{ij}= \overline{B_{ji}}$, and  $ C_{ij}= \overline{C_{ji}}$,
        \item $\Phi$ is $\rm CP$ if and only if $A_{ij} \geq 0$, the matrix $\mathbb{B}$ defined by  $\mathbb{B}_{ii}= A_{ii}$ and 
        $\mathbb{B}_{ij}= B_{ij}$ is positive definite, and $|C_{ij}|^2 \leq A_{ij} A_{ji}$. 
\end{itemize}

\begin{theorem} \label{def:bipartite_orthogonal_covariant}
   A supermap $\Theta$ is diagonal orthogonal covariant if its representing map  $\Delta$   has the following canonical form
   \begin{align}
       \Delta  = 
\Delta_A+ (\Delta_B+\Delta_R)
+ (\Delta_C+\Delta_E)
+ (\Delta_D+\Delta_P+\Delta_Q+\Delta_S).
%\sum_{k=1}^9 \Delta_k , 
  \label{equ:bipartite_orthogonal_covariant}
   \end{align}
where $\{\Delta_A=\Delta _1,\Delta_B=\Delta_2,\Delta_C=\Delta_3,\Delta_D=\Delta_4\}$ are defined in Eq. \eqref{Delta-1} and 
\begin{align}\label{D3}
\Delta_E\left(X\right) &:= \sum_{i\neq j, a b} E_{ia,jb} \left(e_{ii} \otimes e_{ab} \right) X^{\Gamma_1} \left(e_{jj} \otimes e_{ab} \right),  \nonumber \\
\Delta_P\left(X\right)& := 
  \sum_{i\neq j, a \neq b} P_{ia,jb} \left(e_{ii} \otimes e_{aa} \right) X^{\Gamma_1}\left(e_{ii} \otimes e_{bb} \right), \nonumber\\
  \Delta_Q\left(X\right)&:=\sum_{i\neq j, a \neq b} Q_{ia,jb} \left(e_{ii} \otimes e_{aa} \right) X^T \left(e_{jj} \otimes e_{bb} \right),\\
     \Delta_R\left(X\right)& := \sum_{ij, a \neq b} R_{ia,jb} \left(e_{ij} \otimes e_{aa} \right) X^{\Gamma_2} \left(e_{ji} \otimes e_{bb} \right), \nonumber\\
   \Delta_S\left(X\right)&:= 
  \sum_{i\neq j, a\neq b} S_{ia,jb} \left(e_{ii} \otimes e_{aa} \right) X^{\Gamma_2} \left(e_{jj} \otimes e_{bb} \right), \nonumber    
\end{align}
where $X^{\Gamma_{1(2)}}$ is the partial transpose with respect to subsystem $1$ or $2$, that is, 

\[   X^{\Gamma_1} = (T_{A_0} \otimes {\rm id}_{A_1})X \ , \ \ \  X^{\Gamma_2} = ({\rm id}_{A_0} \otimes T_{A_1})X ,\]
where $T_{A_0}$ and $T_{A_1}$ denote transposition maps on $\mathcal{B}(\mathcal{H}^{A_0})$ and $\mathcal{B}(\mathcal{H}^{A_1})$, respectively. 
\end{theorem}
{
\begin{proof}

Let
\[
O_1=\operatorname{diag}(\varepsilon_1,\ldots,\varepsilon_d),
\qquad
O_2=\operatorname{diag}(\eta_1,\ldots,\eta_d),
\]
where \(\varepsilon_i,\eta_a\in\{\pm1\}\). The induced action on
\(\Md \otimes \Md\) is
\[
[\mathcal{O}_1 \otimes \mathcal{O}_2](X)
:=(O_1\otimes O_2)X(O_1\otimes O_2)^T,
\]
and
\[
[\mathcal{O}_1 \otimes \mathcal{O}_2](e_{ij}\otimes e_{ab})
=
\varepsilon_i\varepsilon_j\eta_a\eta_b\,
(e_{ij}\otimes e_{ab}).
\]
Unlike the diagonal unitary case, the weights satisfy now
\[
\varepsilon_i\varepsilon_j
=
\varepsilon_j\varepsilon_i,
\qquad
\eta_a\eta_b
=
\eta_b\eta_a,
\]
so that the matrix units \(e_{ij}\) and \(e_{ji}\) belong to the same
weight space. Consequently, the representation decomposes into the
following weight space:
\[
\mathcal V_{0;0}
=
\operatorname{span}\{e_{ii}\otimes e_{aa}:1\le i,a\le d\},
\]
\[
\mathcal V_{0;{a,b}}
=
\operatorname{span}
\{e_{ii}\otimes e_{ab},
e_{ii}\otimes e_{ba}:1\le i\le d\},
\qquad a<b,
\]
\[
\mathcal V_{{i,j};0}
=
\operatorname{span}
\{e_{ij}\otimes e_{aa},
e_{ji}\otimes e_{aa}:1\le a\le d\},
\qquad i<j,
\]
and
\[
\begin{aligned}
\mathcal V_{{i,j};{a,b}}
=
\operatorname{span}\{
e_{ij}\otimes e_{ab},
e_{ij}\otimes e_{ba},
e_{ji}\otimes e_{ab},
e_{ji}\otimes e_{ba}
\},
\end{aligned}
\qquad i<j,\; a<b.
\]
Since a covariant supermap is an intertwiner of this representation,
it preserves each of the above weight components. Consequently,
the one-dimensional sectors of the diagonal unitary case are replaced
by two- and four-dimensional sectors, giving rise to the nine
independent components
\[
\Delta
=
\Delta_A+ (\Delta_B+\Delta_R)
+ (\Delta_C+\Delta_E)
+ (\Delta_D+\Delta_P+\Delta_Q+\Delta_S).
\]
Explicitly,
\[
\Delta_A:
e_{jj}\otimes e_{bb}
\longmapsto
\sum_{i,a}
A_{ia,jb}\,
e_{ii}\otimes e_{aa},
\]

\[
\Delta_B:
e_{jj}\otimes e_{ab}
\longmapsto
\sum_i
B_{ia,jb}\,
e_{ii}\otimes e_{ab},
\qquad a\neq b,
\]

\[
\Delta_R:
e_{jj}\otimes e_{ab}
\longmapsto
\sum_i
R_{ia,jb}\,
e_{ii}\otimes e_{ba},
\qquad a\neq b,
\]

\[
\Delta_C:
e_{ij}\otimes e_{bb}
\longmapsto
\sum_a
C_{ia,jb}\,
e_{ij}\otimes e_{aa},
\qquad i\neq j,
\]

\[
\Delta_E:
e_{ij}\otimes e_{bb}
\longmapsto
\sum_a
E_{ia,jb}\,
e_{ji}\otimes e_{aa},
\qquad i\neq j,
\]
and, for \(i\neq j\) and \(a\neq b\),
\[
\Delta_D:
e_{ij}\otimes e_{ab}
\longmapsto
D_{ia,jb}\,
e_{ij}\otimes e_{ab},
\]

\[
\Delta_P:
e_{ij}\otimes e_{ab}
\longmapsto
P_{ia,jb}\,
e_{ij}\otimes e_{ba},
\]

\[
\Delta_Q:
e_{ij}\otimes e_{ab}
\longmapsto
Q_{ia,jb}\,
e_{ji}\otimes e_{ab},
\]

\[
\Delta_S:
e_{ij}\otimes e_{ab}
\longmapsto
S_{ia,jb}\,
e_{ji}\otimes e_{ba}.
\]
A simple algebra shows that the above maps coincide with (\ref{D3}).     
\end{proof}
}
The corresponding Choi matrix has the following form

\begin{align}
C_\Delta&= \sum_{ia,jb} A_{ia,jb} \left( e_{ii} \otimes e_{aa} \otimes e_{jj} \otimes e_{bb}\right)+  \sum_{i j,a \neq b} B_{ia,jb} \left( e_{ii} \otimes e_{ab} \otimes e_{jj} \otimes e_{ab}\right) \nonumber\\
&+  \sum_{i\neq j,a,b} C_{ia,jb} \left( e_{ij} \otimes e_{aa} \otimes e_{ij} \otimes e_{bb}\right)+\sum_{i\neq j,a\neq b}  D_{ia,jb} \left( e_{ij} \otimes e_{ab} \otimes e_{ij} \otimes e_{ab}\right)\nonumber\\
&
+ \sum_{i\neq j,a,b} E_{ia,jb} \left( e_{ij} \otimes e_{aa} \otimes e_{ji} \otimes e_{bb}\right)+ \sum_{i\neq j,a \neq b} P_{ia,jb} \left( e_{ij} \otimes e_{ab} \otimes e_{ji} \otimes e_{ab}\right)\nonumber\\
&+ \sum_{i\neq j,a\neq b} Q_{ia,jb} \left( e_{ij} \otimes e_{ab} \otimes e_{ji} \otimes e_{ba}\right)+ \sum_{i j,a\neq b} R_{ia,jb} \left( e_{ii} \otimes e_{ab} \otimes e_{jj} \otimes e_{ba}\right)\nonumber\\
&+\sum_{i\neq j,a\neq b} S_{ia,jb} \left( e_{ij} \otimes e_{ab} \otimes e_{ij} \otimes e_{ba}\right).
\end{align}
In particular for $d=2$, the Choi matrix $C_{\Delta}$ reads as follows (compare with (\ref{C-d=2}))
\begin{equation}   \label{C_DO}
C_\Delta =    \left(
\begin{array}{cccc|cccc|cccc|cccc}
\begin{array}{cccc|cccc|cccc|cccc}
A & . & . & . & . & B & . & . & . & . & C & . & . & . & . & D \\
. & A & . & . & R & . & . & . & . & . & . & C & . & . & S & . \\
. & . & A & . & . & . & . & B & E & . & . & . & . &  P &. & . \\
. & . & . & A & . & . & R & . & . & E & . & . & Q & . & . & . \\\hline
. & R & . & . & A & . & . & . & . & . & . & S & . & . & C & . \\
B & . & . & . & . & A & . & . & . & . & D & . & . & . & . & C \\
. & . & . & R & . & . & A & . & . & Q & . & . & E & . & .& . \\
. & . & B & . & . & . & . & A & P & . & . & . & . & E & . & . \\\hline
. & . & E & . & . & . & . & P & A & . & . & . & . & B & . & . \\
. & . & . & E & . & . & Q & . & . & A & . & . & R & . & . & . \\
C & . & . & . & . & D & . & . & . & . & A & . & . & . & . & B \\
. & C & . & . & S & . & . & . & . & . & . & A & . & . & R & . \\\hline
. & . & . & Q & . & . & E & . & . & R & . & . & A & . & . & . \\
. & . & P & . & . & . & . & E & B & . & . & . & . & A & . & . \\
. & S & . & . & C & . & . & . & . & . & . & R & . & . & A & . \\
D & . & . & . & . & C & . & . & . & . & B & . & . & . & . & A
\end{array}
\end{array} \right) .
\end{equation}
{

The above decomposition describes the most general diagonal-orthogonal
covariant linear supermap. To obtain a quantum superchannel one must still
impose the causality, or trace-preservation constraint
\[
\operatorname{Tr}_{B_1}\circ \Delta
=
\widetilde{\Delta}\circ \operatorname{Tr}_{A_1},
\]
for some linear map  $\widetilde{\Delta}: \Md\to \Md$
satisfying $\widetilde{\Delta}(\oper)=\oper$. 
Equivalently, for all \(X\in M_d\otimes M_d\), $\operatorname{Tr}_{B_1}\Delta(X)$ 
must depend only on \(\operatorname{Tr}_{A_1}X\), and not on the individual
blocks of \(X\). The causality constraint is therefore imposed only on the
components which produce diagonal blocks, namely the \(A,C,E\)-type
terms. More explicitly, the coefficients must satisfy
\[
\sum_a A_{ia,jb}=\alpha_{ij},
\qquad
\sum_a C_{ia,jb}=\gamma_{ij},
\qquad
\sum_a E_{ia,jb}=\eta_{ij},
\]
independently of \(b\). The reduced map is DO-covariant and reads as follows
\[
\widetilde{\Delta}(X)
=
\sum_{i,j}
\alpha_{ij}\,e_{ij}Xe_{ji}
+
\sum_{i\neq j}
\gamma_{ij}\,e_{ii}Xe_{jj}
+
\sum_{i\neq j}
\eta_{ij}\,e_{ii}X^T e_{jj}.
\]
Finally, the normalization condition $\widetilde{\Delta}(\oper)=\oper$ implies
\[
\sum_j \alpha_{ij}=1,
\qquad i=1,\ldots,d.
\]
Thus, the covariance decomposition determines the possible block structure,
whereas the causality constraint selects those covariant maps that are
admissible one-slot quantum combs, i.e. genuine quantum superchannels.

One easily computes the number of independent complex parameters:
\[
A:\quad d^4,
\]
\[
B,C,E,R:\quad d^2\,d(d-1)=d^3(d-1),
\]
and
\[
D,P,Q,S:\quad d(d-1)\,d(d-1)=d^2(d-1)^2.
\]
Hence, the total number of independent complex parameters is
\[
\begin{aligned}
N_{\mathrm{DO}}
=
d^4
+
4d^3(d-1)
+
4d^2(d-1)^2 
=
d^2(3d-2)^2.
\end{aligned}
\]
A similar count of parameters applies to the Hermiticity preserving maps. In this case, one has $N_{\rm DO}$ real parameters. Adding the TP preservation constraints yields

\[ N_{\mathrm{DO}}^{\rm TP} = 3d(d-1)(3d^2-2d +1) .  \]

\begin{remark}
Note that for $d=2$, one has $N_{\rm DO}=64$ which perfectly agrees with the total number of nonvanishing matrix elements in (\ref{C_DO}).  Interestingly, the number of independent complex parameters of any DO-covariant map $\Md\to \Md$ equals to $n_{\rm DO}= d^2 + d(d-1) +d(d-1) = d(3d-2)$ and hence $N_{\rm DO}=n_{\rm DO}^2$. 
\end{remark}

\begin{proposition} \label{prop_ocp}
A diagonal-orthogonal covariant map \(\Delta\) is completely positive iff
\[
K_0\geq 0
\]
and
\[
K_{ij}\geq 0,
\qquad 1\leq i<j\leq d,
\]
where
\[
K_0
=
\sum_i e_{ii} \otimes e_{ii} \otimes M_{ii}
+
\sum_{i\neq j} e_{ij} \otimes e_{ij} \otimes N_{ij},
\]
and
\[
K_{ij}
=
\begin{pmatrix}
M_{ij} & L_{ij}\\
L_{ji} & M_{ji}
\end{pmatrix}.
\]
Here \(M_{ij},N_{ij},L_{ij}\in B(\mathcal H^{A_1B_1})\) are defined by
\[
M_{ij}
=
\sum_{a,b}
A_{ja,ib}\,e_{bb}\otimes e_{aa}
+
\sum_{a\neq b}
B_{ja,ib}\,e_{ab}\otimes e_{ab}
+
\sum_{a\neq b}
R_{ja,ib}\,e_{ab}\otimes e_{ba},
\]

$$N_{ij}
=
\sum_{a,b}
C_{ia,jb}\,e_{aa}\otimes e_{bb}
+
\sum_{a\neq b}
D_{ia,jb}\,e_{ab}\otimes e_{ab}
+
\sum_{a\neq b}
S_{ia,jb}\,e_{ab}\otimes e_{ba},$$
and

$$L_{ij}
=
\sum_{a,b}
E_{ia,jb}\,e_{aa}\otimes e_{bb}
+
\sum_{a\neq b}
P_{ia,jb}\,e_{ab}\otimes e_{ab}
+
\sum_{a\neq b}
Q_{ia,jb}\,e_{ab}\otimes e_{ba}.$$
\end{proposition}
\begin{proof}
The map \(\Delta\) is completely positive if and only if its Choi
matrix \(C_\Delta\) is positive semidefinite. Since the flip
\(P= \oper_{A_0}\otimes F_{A_1B_0}\otimes \oper_{B_1}\)
is unitary, this is equivalent to positivity of
\[
C_\Delta^F:=PC_\Delta P^\dagger .
\]
Using the diagonal-orthogonal covariant decomposition of \(\Delta\),
one obtains
\[
C_\Delta^F
=
K_0
+
\sum_{i<j}
K_{ij},
\]
where the summands have mutually orthogonal supports. More precisely,
the first term acts on
\[
\mathcal W_0
=
\operatorname{span}_{\mathbb{C}}\{ e_i \otimes e_i\ :\ 1\leq i\leq d\}
\otimes \mathcal H^{A_1B_1},
\]
whereas, for each \(i<j\), \(K_{ij}\) acts on
\[
\mathcal W_{ij}
=
\operatorname{span}_{\mathbb{C}}\{\, e_i \otimes e_j , e_j \otimes e_i\}
\otimes \mathcal H^{A_1B_1}.
\]
Thus
\[
\mathcal H^{A_0B_0A_1B_1}
=
\mathcal W_0
\oplus
\bigoplus_{i<j}\mathcal W_{ij},
\]
and this is an invariant subspace decomposition for \(C_\Delta^F\).
On \(\mathcal W_0\), the restriction of \(C_\Delta^F\) is
\[
K_0
=
\sum_i e_{ii} \otimes e_{ii} \otimes M_{ii}
+
\sum_{i\neq j} e_{ij} \otimes e_{ij}\otimes N_{ij}.
\]
On \(\mathcal W_{ij}\), using the ordered basis
\[
e_i \otimes e_j \otimes \mathcal H^{A_1B_1},
\qquad
e_j \otimes e_i \otimes \mathcal H^{A_1B_1},
\]
the restriction is the \(2\times2\) operator-valued block matrix
\[
K_{ij}
=
\begin{pmatrix}
M_{ij} & L_{ij}\\
L_{ji} & M_{ji}
\end{pmatrix}.
\]
Consequently,
\[
C_\Delta^F
\simeq
K_0
\oplus
\bigoplus_{i<j}K_{ij}.
\]
Therefore
\[
C_\Delta^F\geq0
\quad\Longleftrightarrow\quad
K_0\geq0
\ \text{and}\
K_{ij}\geq0
\quad\text{for all }i<j.
\]
Since \(C_\Delta\geq0\) if and only if \(C_\Delta^F\geq0\), the claim follows.
\end{proof}

\begin{proposition}
Let \(\widetilde{\Delta}=\Delta'\circ\Delta\). The corresponding coefficients
\((\widetilde A,\widetilde B,\widetilde R,\widetilde C,\widetilde E,
\widetilde D,\widetilde P,\widetilde Q,\widetilde S)\)
are given as follows. \\
For the $A$-sector:
\[
\widetilde A_{ia,kc}
=
\sum_{j,b}
A'_{ia,jb}\,A_{jb,kc}.
\]
For the \((B,R)\)-sector:

\[  \begin{pmatrix}
    \widetilde B \\ \widetilde R 
\end{pmatrix}_{ia,kb} = \sum_j \begin{pmatrix}
    B'_{ia,jb} & R'_{ib,ja}  \\ R'_{ia,jb} & B'_{ib,ja} 
\end{pmatrix} \begin{pmatrix}
     B \\ R 
\end{pmatrix}_{ja,kb}\ , \quad a \neq b . \]
For the \((C,E)\)-sector:

\[  \begin{pmatrix}
    \widetilde C \\ \widetilde E 
\end{pmatrix}_{ia,jc} = \sum_b \begin{pmatrix}
    C'_{ia,jb} & E'_{ja,ib}  \\ E'_{ia,jb} & E'_{ja,ib} 
\end{pmatrix} \begin{pmatrix}
     C \\ E 
\end{pmatrix}_{ib,jc} \ , \quad i \neq j .\]
Finally, for the $(D,P,Q,S)$-sector: %(i.e. \(i\neq j\) and \(a\neq b\)):

\[   \begin{pmatrix}
    \widetilde{D} \\  \widetilde{P} \\  \widetilde{Q} \\  \widetilde{D}  
\end{pmatrix}_{ia,jb} = 
\begin{pmatrix}
    D'_{ia,jb} & P'_{ib,ja} & Q'_{ja,ib}& S'_{jb,ia} \\
    P'_{ia,jb} & D'_{ib,ja} & S'_{ja,ib} & Q'_{jb,ia} \\
    Q'_{ia,jb} & S'_{ib,ja} & D'_{ja,ib} & P'_{jb,ia} \\
    S'_{ia,jb} & Q'_{ib,ja} & P'_{ja,ib} & D'_{jb,ia} 
\end{pmatrix} \begin{pmatrix}
    D \\ P \\ Q \\ S
\end{pmatrix}_{ia,jb} \ ,  \quad i \neq j \ , \ a \neq b .
\]
    
\end{proposition}
}

Diagonal orthogonal covariant supermaps are stable under the composition of supermaps. The corresponding multiplication rules $\Delta_i \circ \Delta'_j$ can easily be obtained but we do not present them here. Interestingly, (\ref{C_DO}) has a circulant structure exposed in \cite{circulant_states}.

\section{The Action of Diagonal Unitary Covariant Maps}\label{sec6} 

Let $X = \sum_{i,j} e_{ij} \otimes X_{ij} $, where $X_{ij} \in \Md$. Then following (\ref{Delta-1}) it is easy to find 

\begin{eqnarray}
    \Delta_1(X) &=& \sum_{i,j} e_{ii} \otimes \Big( \sum_{a,b} A_{ia,jb}\  e_{ab}\, X_{jj}  \, e_{ba} \Big) \nonumber\ , \\
     \Delta_2(X) &=& \sum_{i,j} e_{ii} \otimes \Big(\sum_{a\neq b} B_{ia,jb}\ e_{aa}\, X_{jj} \,  e_{bb} \Big) \ , \\
       \Delta_3(X) &=& \sum_{i\neq j} e_{ij} \otimes \Big( \sum_{a,b} C_{ia,jb}\ e_{ab}\,  X_{ij} \,  e_{ba} \Big) \ , \nonumber\\
        \Delta_4(X) &=& \sum_{i\neq j} e_{ij} \otimes \Big( \sum_{a\neq b} D_{ia,jb}\ e_{aa}\, X_{ij} \,  e_{bb} \Big) \nonumber\ .
\end{eqnarray}
Hence, $\Delta_1$ and $\Delta_2$ affect only the diagonal blocks of $X$, whereas $\Delta_3$ and $\Delta_4$ affect only the off-diagonal blocks of $X$. Moreover, $\Delta_1$ affects only the diagonal elements of $X_{jj}$, whereas $\Delta_2$ affects only the off-diagonal elements of $X_{jj}$. Similarly, $\Delta_3$ affects only the diagonal elements of $X_{ij}$ $(i\neq j)$, whereas $\Delta_4$ affects only the off-diagonal elements of $X_{ij}$ $(i\neq j)$. 

Let $Y = \Delta(X) = \sum_{i,j} e_{ij} \otimes Y_{ij}$. This yields for the diagonal blocks $Y_{ii}$ 

\begin{eqnarray*}
    (Y_{ii})_{aa} &=& \sum_{j,b} A_{ia,jb} \ (X_{jj})_{bb} \ , \\
     (Y_{ii})_{ab} &=& \sum_{j} B_{ia,jb} \ (X_{jj})_{ab} \ , \ \ \ a \neq b ,
\end{eqnarray*}
and for the off-diagonal blocks $Y_{ij}$ $(i\neq j)$:    
\begin{eqnarray*}
      (Y_{ij})_{aa} &=& \sum_{b} C_{ia,jb} \ (X_{ij})_{bb} \ ,\ \ \ i \neq j \\
       (Y_{ij})_{ab} &=& D_{ia,jb} \ (X_{ij})_{ab} \ , \ \ \ i \neq j \ , \ a \neq b . 
\end{eqnarray*}
{Table \ref{tab:ABCD} explains the role of the sectors $A, B, C$ and $D$.

\begin{table}[t]
\centering
{
\begin{tabular}{|c|l|l|}
\hline
\textbf{Sector} & \textbf{Input quantity} & \textbf{Output quantity} \\
\hline
$A$ &
Populations of diagonal blocks &
Populations of diagonal blocks \\
\hline
$B$ &
Coherences in diagonal blocks &
Coherences in diagonal blocks \\
\hline
$C$ &
Populations in off-diagonal blocks &
Populations in off-diagonal blocks \\
\hline
$D$ &
Coherences in off-diagonal blocks &
Coherences in off-diagonal blocks \\
\hline
\end{tabular}
\caption{Interpretation of the four sectors in the canonical decomposition of a diagonal unitary covariant superchannel.}
\label{tab:ABCD}
}
\end{table}

}

\begin{Example}
For $d=2$, the above general formulae give rise to the following four $2 \times 2$ blocks $Y_{ij}$ --- the diagonal blocks:

\begin{equation}
    Y_{11} =  \left( \begin{array}{cc} 
   \sum_{j,b} A_{11,jb} (X_{jj})_{bb}  & \sum_j B_{11,j2} \ (X_{jj})_{12} \\
\sum_j B_{12,j1} \ (X_{jj})_{21} & \sum_{j,b} A_{12,jb} (X_{jj})_{bb}
\end{array} \right) ,
\end{equation}

\begin{equation}
    Y_{22} =  \left( \begin{array}{cc} 
   \sum_{j,b} A_{21,jb} (X_{jj})_{bb}  & \sum_j B_{21,j2} \ (X_{jj})_{12} \\
\sum_j B_{22,j1} \ (X_{jj})_{21} & \sum_{j,b} A_{22,jb} (X_{jj})_{bb}
\end{array} \right) ,
\end{equation}
and the off-diagonal blocks: 

\begin{equation}
    Y_{12} =  \left( \begin{array}{cc} \sum_b C_{11,2b} (X_{12})_{bb}
   %C_{11,21} X_{11,21} + C_{11,22} X_{12,22} 
   & D_{11,22} (X_{12})_{12} \\ 
  D_{12,21} (X_{12})_{21} & \sum_b C_{12,2b} (X_{12})_{bb}
 % \sum_b C_{21,1b} (X_{11})_{bb}C_{11,21} X_{11,21} + C_{11,22} X_{12,22}
\end{array} \right) ,
\end{equation}

\begin{equation}
    Y_{21} =  \left( \begin{array}{cc} \sum_b C_{21,1b} (X_{21})_{bb}
  % C_{21,11} X_{21,11} + C_{22,12} X_{22,12} 
  & D_{21,12} (X_{21})_{12} \\ 
   D_{22,11} (X_{21})_{21} & \sum_b C_{22,1b} (X_{21})_{bb}
   %C_{21,11} X_{21,11} + C_{22,12} X_{22,12}
\end{array} \right) .
\end{equation}
\end{Example}
It is therefore clear that, under the action of $\Delta$, the diagonal blocks $X_{ii}$ and the off-diagonal blocks $X_{ij}$ ($i \neq j$) transform independently. This is a direct generalization of the well known property of  diagonal unitary covariant quantum channels, which transform diagonal and off-diagonal elements of density matrices independently. In particular, one has the following proposition.

\begin{proposition}\label{prop12}
    A supermap $\Theta$ that is $\rm DU$--covariant, maps the diagonal orthogonal covariant maps into the diagonal orthogonal covariant maps.
\end{proposition}
Indeed, the Choi matrix of a diagonal orthogonally covariant map is given by $ X = X_{1}+X_{2}+X_{3}$, where 
$$ X_{1}=\sum_{m,n} P_{mn} \, e_{mm} \otimes e_{nn}\ , \ \ X_{2}=\sum_{m\neq n} Q_{mn}\, e_{mn} \otimes e_{mn} \ , \ \ X_{3}=\sum_{m\neq n} R_{mn}\, e_{mn} \otimes e_{nm}\ . $$
The action of $\Delta_\Theta$ on $X$ generates total $12$ terms, but a simple computation shows that only three of them are non-zero, that is, 
\begin{align}
    \Delta_{1} \left(X_1\right)& = \sum_{ij,ab} A_{ia,jb}\, P_{jb}\, e_{ii}\otimes e_{aa}, \nonumber\\
    \Delta_{4} \left(X_2\right) &= \sum_{i\neq j} D_{ii,jj}\, Q_{ij}\, e_{ij}\otimes e_{ij},\\
    \Delta_{4} \left(X_3\right) &= \sum_{i\neq j} D_{ij,ji}\, R_{ij}\, e_{ij}\otimes e_{ji} \nonumber.
\end{align}
Clearly, these terms are invariant under a diagonal orthogonal group, and hence,

$$  \Delta(X_1+X_2+X_3) = \Delta_1(X_1) + \Delta_4(X_2+X_3)$$
is diagonal orthogonally invariant. In particular for $d=2$,

\begin{equation}
    \Delta \left( \begin{array}{cc|cc} a_1 & . & . & \mu \\ . & a_2 &  \nu & . \\ \hline
    . & \overline{\nu} & a_3 & . \\ \overline{\mu} & . & . & a_4 \end{array} \right) = 
    \left( \begin{array}{cc|cc} a'_1 & . & . & \mu' \\ . & a'_2 &  \nu' & . \\ \hline
    . & \overline{\nu}' & a'_3 & . \\ \overline{\mu}' & . & . & a'_4 \end{array} \right).
\end{equation}
Finally, let us observe how diagonal unitary covariant superchannels act on an identity channel. {At} the level of a representing map $\Delta_\Theta$ one finds

\begin{equation}
    \Delta_\Theta(P^+) = \sum_{i,j} S_{ij}\, e_{jj} \otimes e_{ii} + \sum_{i\neq j} D_{ii,jj}\,e_{ij} \otimes e_{ij} \ ,
\end{equation}
with

\begin{equation}
    S_{ij} = \sum_k A_{ji,kk} .
\end{equation}
Note that $S_{ij}$ defines a column stochastic matrix. Indeed, one has 

$$  \sum_i S_{ij} = \sum_k \Big( \sum_i A_{ji,kk} \Big) = \sum_k \alpha_{jk} = 1 ,$$
due to the fact that $\alpha_{jk} = \sum_i A_{ji,k\ell}$ is row stochastic for any $\ell=1,\ldots,d$. 
Equivalently,

\begin{equation}
\Theta[{\rm id}](X) = \sum_{i,j} S_{ij}\, e_{ij} X e_{ji} + \sum_{i\neq j} D_{ii,jj}\, e_{ii} X e_{jj} .
\end{equation}

\begin{corollary} Diagonal unitary covariant quantum superchannels acting on the identity channel generate a class of diagonal unitary covariant quantum channels.
\end{corollary}

\section{Dephasing Superchannels}\label{sec7}

Having established the general structure of diagonal unitary covariant superchannels, we now focus on an important subclass of dephasing superchannels \cite{Korzekwa2018,Puchala2021}. Recall that a dephasing quantum channel is defined in terms of the Schur (Hadamard) product

\begin{equation} \label{Phi-M}
    \Phi(X) = M \odot X ,
\end{equation}
where $M$ is a covariance matrix, i.e. $M \geq 0$ and $M_{ii}=1$. It is evident that such channels are DU-covariant, {with $A_{ij} = \delta_{ij}$, that is, }

\begin{equation}
    M_{ij} = \left\{ \begin{array}{ll} B_{ij} & ; \ i\neq j \\ A_{ii}=1 & ; \ i=j \end{array} \right. .
\end{equation}
Dephasing quantum channels may be realized as

\begin{equation}  \label{Phi-UU}
    \Phi(X) = {\rm Tr}_E \Big( U\, (X \otimes |\psi_E\>\<\psi_E|)\, U^\dagger \Big) , 
\end{equation}
where the unitary $U$ is block diagonal, i.e., $U = \sum_i e_{ii} \otimes U_i$, with $U_i$ being unitaries on the Hilbert space of the environment, and $|\psi_E\> \in \mathcal{H}^E$ is an arbitrary (fixed) state vector. One derives $\Phi(X) = M \odot X$, where $M$ is a correlation matrix defined by
$$   M_{ij} = \<\psi_E|U_j^\dagger U_i| \psi_E\> . $$
This construction may be immediately generalized to quantum superchannels \cite{Puchala2021}.

{
\begin{definition}
A superchannel $\Theta$ is  called dephasing (w.r.t. the distinguished basis $|i\>$) iff for any linear map $\Phi$ one has 

\begin{equation}
   \< i| \Theta[\Phi](|j\rangle \langle j|)|i\> = \< i|\Phi(|j\>\<j|)|i\> .
\end{equation}
\end{definition}
Equivalently, $\Theta$ is dephasing if its representing map is defined via the Schur product
}

\begin{eqnarray}
    \Delta_\Theta(C_\Phi) = \mathbb{M} \odot C_\Phi , 
\end{eqnarray}
that is,

\begin{eqnarray}
    \Delta_\Theta(C_\Phi) = \sum_{i,j}\sum_{a,b} \mathbb{M}_{ia,jb} \, e_{ii} \otimes e_{aa}\, C_\Phi \, e_{jj} \otimes e_{bb}  \ .
\end{eqnarray}
Such a supermap is DU-covariant, and the matrix $\mathbb{M}$ is defined as follows:

\begin{equation}
    \mathbb{M}_{ia,jb} = \left\{ \begin{array}{ll} D_{ia,jb} & ; \ i\neq j, \ a\neq b  \\
    B_{ia,ib} & ; \ i=j, \ a\neq b  \\
    C_{ia,ja} & ; \ i\neq j, \ a = b  \\
    A_{ia,ia} & ; \ i = j, \ a = b  \end{array} \right. . 
\end{equation}
{ It provides, therefore, the following constraints for the matrices $A,B,C$:

\begin{equation}
    A_{ia,jb} ={A_{ia,ia}} \,\delta_{ij}\delta_{ab} \ , \ \ B_{ia,jb} = \delta_{ij}  B_{ia,ib} \ , \ \ C_{ia,jb} = \delta_{ab}  C_{ia,ja} .
\end{equation}
Interestingly, 

\begin{proposition}
The cone of completely positive Schur-product supermaps forms a face of the cone of
DU-covariant completely positive supermaps.
\end{proposition}

\begin{proof}
A Schur-product supermap is represented by
\[
\Delta_M(X)= \mathbb{M}\odot X,
\]
where $\mathbb{M}=(\mathbb{M}_{ia,jb})$ is positive semidefinite. 
The corresponding Choi matrix
\[
C_{\Delta_M}
=
\sum_{i,j} \sum_{a,b} \mathbb{M}_{ia,jb} \, e_{ij} \otimes e_{ab} \otimes e_{ij} \otimes e_{ab}
\]
is supported on

\[
\mathcal S
=
\operatorname{span}_{\mathbb{C}} \{ e_i \otimes e_a \otimes e_i \otimes e_a\} .
\]
Let $\mathcal C_{\rm DU}$ denote the cone of DU covariant completely
positive supermaps. The Schur-product cone is therefore defined via
\[
\mathcal C_{\rm Schur}
=
\{
C_\Delta\in\mathcal C_{\rm DU} \, | \, \operatorname{supp}(C_\Delta)\subseteq\mathcal S
\}.
\]
Finally, to prove that $\mathcal C_{\rm Schur}$ is a face of $\mathcal C_{\rm DU}$,
suppose that
\[
C=C_1+C_2 \in \mathcal C_{\rm Schur} \]
for some $C_1,C_2\in\mathcal C_{\rm DU}$. Since $C_1,C_2\ge0$, one has $
0\le C_i\le C$ for $i=1,2$, and hence $
\ker C
\subseteq
\ker C_i$. One has therefore

\[
\operatorname{supp}(C_i)
\subseteq {\operatorname{supp}(C)} \subseteq 
\mathcal S, 
\qquad i=1,2.
\]
Hence each $C_i$ is again the Choi matrix of a Schur-product supermap 
which proves that $\mathcal C_{\rm Schur}$ is a face of the cone
$\mathcal C_{\rm DU}$.
\end{proof}

}

Now, $\Delta_\Theta$ is completely positive iff $\mathbb{M} \geq 0$. Moreover, representing $X \in \Md \otimes \Md$ via  $X= \sum_{c,d} X_{cd} \otimes e_{cd}$, one has

\begin{equation}
    {\rm Tr}_{B_1} \Delta_\Theta(X) = \sum_{i,j} \sum_{a} \mathbb{M}_{ia,ja} \, e_{ii} \,X_{aa}\, e_{jj} .
\end{equation}
Hence, $\Theta$ is TP preserving iff $\mathbb{M}_{ia,ja} = M_{ij}$ independently of $a$. Indeed, one has in this case

\begin{equation}
    {\rm Tr}_{B_1} \Delta_\Theta(X) = \sum_{i,j}  {M}_{ij} \, e_{ii} \,\Big( \sum_{a} X_{aa} \Big) \, e_{jj} = \widetilde{\Delta}_\Theta({\rm Tr}_{A_1}X) ,
\end{equation}
where $ \widetilde{\Delta}_\Theta : \Md \to \Md$ reads

\begin{equation}
    \widetilde{\Delta}_\Theta(A) = M \odot A , 
\end{equation}
for any $A \in \Md$. Summarizing, $\Theta$ is TP preserving iff $\mathbb{M}_{ia,ja} = M_{ij}$ and $M_{ii}=1$~\cite{Puchala2021}. 

{
\begin{remark} Note that the TP-preserving condition is perfectly consistent with the general TP-preserving condition for DU-covariant supermaps derived in Proposition \ref{propp6}. Indeed, 

\[ \sum_a A_{ia,jb} = \delta_{ij} \ , \ \ \ \sum_a C_{ia,jb} = \sum_a \delta_{ab} C_{ia,ja} = C_{ib,jb} , \]
and hence $C_{ia,ja}$ does not depend on $a$ which implies     $\mathbb{M}_{ia,ja} = M_{ij}$.
\end{remark}
}

Dephasing superchannels have a natural physical realization (analogous to (\ref{Phi-UU})).

{
\begin{proposition}[\cite{Puchala2021}] Every dephasing superchannel $\Theta$ can be represented as follows:

\begin{equation}
    \Theta[\Phi](X) = {\rm Tr}_E \Big( \Big[\mathcal{V} \circ (\Phi \otimes {\rm id}_E) \circ \mathcal{U} \Big](X \otimes |\psi_E\>\<\psi_E|) \Big)  ,
\end{equation}
where $\mathcal{U}(A) = UAU^\dagger$, $\mathcal{V}(B) = VBV^\dagger$,  and $U,V$ are block diagonal unitaries

$$   U = \sum_i e_{ii} \otimes U_i \ , \quad V = \sum_a e_{aa} \otimes V_a ,$$
with $U_i$ and $V_a$ being unitaries on $\mathcal{H}^E$, where ${\rm dim}\, \mathcal{H}^E \geq {\rm rank}\, \mathbb{M} $. 
\end{proposition}}

One obtains 

\begin{equation}
    \Theta[\Phi](X) = \sum_{i,j}\sum_{a,b} \mathbb{M}_{ia,jb}\, e_{aa} \Phi( e_{ii}X e_{jj}) e_{bb} \ ,
\end{equation}
or, equivalently,

\begin{equation}\\\label{Theta-M}
    \Theta[\Phi] = \sum_{i,j}\sum_{a,b} \mathbb{M}_{ia,jb}\, \mathcal{E}^a_{\ b} \circ  \Phi \circ \mathcal{E}^i_{\ j}  \ , 
\end{equation}
where $\mathcal{E}^i_{\ j}(X) := e_{ii} X e_{jj}$. The matrix $\mathbb{M}$ is defined by

\begin{equation}
    \mathbb{M}_{ia,jb} = \<\psi_E| U_i^\dagger V_b^\dagger V_a U_j|\psi_E\> . 
\end{equation}
Note that, dephasing superchannels transform dephasing channels into dephasing channels. Indeed, if $\Theta$ is a dephasing superchannel defined in (\ref{Theta-M}), and $\Phi$ is a dephasing channel defined in (\ref{Phi-M}), then

\begin{equation}
    \widetilde{\Phi}(X) = \Theta[\Phi](X) = \widetilde{M} \odot X , 
\end{equation}
with $\widetilde{M}_{ij} = \mathbb{M}_{jj,ii}\, M_{ij}$. In particular, acting on an identity channel one obtains

\begin{equation}
    \Theta[{\rm id}] = \sum_{i,j} \mathbb{M}_{ii,jj} \, \mathcal{E}^i_{\ j} ,
\end{equation}
which nicely illustrates the process of {\em {superdecoherence}} on the level of channels instead of the standard decoherence on the level of states \cite{Korzekwa2018,Rico2025}.

\section{Examples: Transforming Qubit Channels}\label{sec8} 

In this section, we illustrate the presented formalism by the action of {{\rm DU}-covariant } superchannels on well known qubit quantum channels. 

\begin{Example}[Amplitude-damping channel]
{\em     
The qubit amplitude-damping channel $\Phi$ with damping parameter
$\gamma \in [0,1]$ is defined in the Kraus form by
\begin{equation}
K_0 =
\begin{pmatrix}
1 & 0\\[2pt]
0 & \sqrt{1-\gamma}
\end{pmatrix},
\qquad
K_1 =
\begin{pmatrix}
0 & \sqrt{\gamma}\\[2pt]
0 & 0
\end{pmatrix},
\end{equation}
so that for any qubit state \(\rho\), 
\begin{equation}  \label{AD}
\Phi(\rho) = K_0 \rho K_0^\dagger + K_1 \rho K_1^\dagger  
    = \sum_{i,j=1}^2 \alpha_{ij}\, e_{ij} \rho e_{ji} + \sqrt{1-\gamma} \Big( e_{11} \rho e_{22} +  e_{22} \rho e_{11} \Big) , 
\end{equation}
where 

$$ \alpha_{ij} =   \begin{pmatrix}  1 &  \gamma \\  0 &  1-\gamma    \end{pmatrix} , $$
is a column-stochastic matrix (a classical channel). The Choi matrix of $\Phi$ reads 

\begin{equation}
C_\Phi =
\begin{pmatrix}
1 & 0 & 0 & \sqrt{1-\gamma}\\[4pt]
0 & 0 & 0 & 0\\[4pt]
0 & 0 & \gamma & 0\\[4pt]
\sqrt{1-\gamma} & 0 & 0 & 1-\gamma
\end{pmatrix}.
\end{equation}
Hence, {the action of a $\rm DU$-covariant superchannel $\Delta_\Theta$ on $C_{\Phi}$ is given by}

\begin{equation}
    \Delta_{\Theta}\left(C_\Phi\right) = \begin{pmatrix}
a_1 & 0 & 0 & D_{11,22}\, \sqrt{1-\gamma}\\[4pt]
0 & a_2 & 0 & 0\\[4pt]
0 & 0 & a_3 & 0\\[4pt]
D_{22,11}\, \sqrt{1-\gamma} & 0 & 0 & a_4 
\end{pmatrix} ,
\end{equation}
with

\begin{eqnarray*}
    a_1 &=& A_{11,11} + A_{11,21} \gamma + A_{11,22} (1-\gamma) , \\
    a_2 &=& A_{12,11} + A_{12,21} \gamma + A_{12,22} (1-\gamma) , \\
    a_3 &=& A_{21,11} + A_{21,21} \gamma + A_{21,22} (1-\gamma) , \\
    a_4 &=& A_{22,11} + A_{22,21} \gamma + A_{22,22} (1-\gamma) .
\end{eqnarray*}
Note that,

\begin{eqnarray}
    a_1+a_2 = \alpha_{11} + \alpha_{12} = 1 , \quad 
    a_3+a_4 = \alpha_{21} + \alpha_{22}   =1 ,
\end{eqnarray}
where $\alpha_{ij} := \sum_a A_{ia,jb}$. Hence, the matrix

$$ a_{ij} =  \begin{pmatrix}  a_1 &  a_3 \\  a_2 &  a_4  \end{pmatrix} $$
is row-stochastic, which proves that  

\begin{equation}
    \Delta_{\Theta}\left(C_\Phi\right) = \begin{pmatrix}
a_1 & 0 & 0 & D_{11,22}\, \sqrt{1-\gamma}\\[4pt]
0 & 1-a_1 & 0 & 0\\[4pt]
0 & 0 & 1-a_4 & 0\\[4pt]
D_{22,11}\, \sqrt{1-\gamma} & 0 & 0 & a_4 
\end{pmatrix} ,
\end{equation}
indeed defines the Choi matrix of a quantum channel. {Consequently,} an amplitude-damping channel (\ref{AD}) is mapped to $\Phi'=\Theta[\Phi]$,

\begin{equation}
    \Phi'(\rho) = \sum_{i,j=1}^2 a_{ij}\, e_{ij} \rho e_{ji} + \sqrt{1-\gamma} \Big( u\, e_{11} \rho e_{22} + \overline{u}\, e_{22} \rho e_{11} \Big) \ , 
\end{equation}
with $u := D_{11,22}$.
}
{Any DU-covariant superchannel maps an amplitude damping qubit channel to DU-covariant qubit channel. Let us observe that there is a subclass of DU-covariant superchannels  which maps any amplitude damping  channel to an amplitude damping  channel. This class is defined via

\begin{equation}
    A(\lambda) = \begin{pmatrix}
1 & 0 & 0 & 0 \\
0 & 1 & 0 & 0 \\
0 & 0 & 1 & 1-\lambda \\
0 & 0 & 0 & \lambda
\end{pmatrix}, \ \ \ \   B(\lambda)= \sqrt{\lambda} 
\begin{pmatrix}
0&1&0&1\\
1&0&1&0\\
0&1&0&1\\
1&0&1&0
\end{pmatrix},
%D(\lambda) = \begin{pmatrix}
%0 & 0 & 0 & \sqrt{\lambda} \\
%0 & 0 & 0 & 0 \\
%0& 0 & 0 & 0 \\
%\sqrt{\lambda} & 0 & 0 & 0
%\end{pmatrix} \ ,
\end{equation}
and 
\begin{equation}
C(\lambda) = \sqrt{\lambda} 
\begin{pmatrix}
0&0&1&1\\
0&0&1&1\\
1&1&0&0\\
1&1&0&0
\end{pmatrix},
\qquad
D(\lambda)= \sqrt{\lambda} 
\begin{pmatrix}
0&0&0&1\\
0&0&1&0\\
0&1&0&0\\
1&0&0&0
\end{pmatrix}. 
\end{equation}
Note that for $\lambda=1$ this defines an idenity superchannel ${\rm Id}$. Moreover,

\[   \Delta(\lambda) \circ \Delta(\lambda') = \Delta(\lambda \lambda') .\]
An amplitude damping channel with the parameter $\gamma$ is transformed into the amplitude damping channel with the parameter $\gamma(\lambda)$
\begin{equation}
   \gamma \rightarrow \gamma(\lambda) := 1-\lambda(1-\gamma).
\end{equation}  
Since $\lambda \in [0,1]$  this transformation  always increases the damping parameter $\gamma(\lambda) \geq \gamma$. 
We propose to call this family an {\em amplitude damping superchannel}.}

\end{Example}

\begin{Example}[Bit-flip channel]
{\em  

The bit-flip channel with flip probability $p\in[0,1]$ acts as
\[
\Phi(\rho)=(1-p)\,\rho + p\,\sigma_x\rho \sigma_x .
\]
A Kraus representation is {obtained from} $K_0=\sqrt{1-p}\,\oper_2$ and $K_1=\sqrt{p}\,\sigma_x$.
The  Choi matrix reads

\[
C_\Phi=
\begin{pmatrix}
1-p & 0 & 0 & 1-p\\[4pt]
0 & p & p & 0\\[4pt]
0 & p & p & 0\\[4pt]
1-p & 0 & 0 & 1-p
\end{pmatrix},
\]
and hence, 

\begin{equation}
    \Delta_\Theta(C_\Phi) =
\begin{pmatrix}
p_{11} & 0 & 0 & D_{11,22}\,(1-p)\\[4pt]
0 & p_{12} & D_{12,21}\,p & 0\\[4pt]
0 & D_{21,12}\, p & p_{21} & 0\\[4pt]
D_{22,11}\,(1-p) & 0 & 0 & p_{22} \label{dubit_flip_choi}
\end{pmatrix} ,
\end{equation}  
where

$$   p_{ij} = p \Big( A_{ij,12} + A_{ij,21} \Big) + (1-p) \Big( A_{ij,11} + A_{ij,22} \Big) . $$
}
{Hence, a DU-covariant superchannel maps a bit-flip channel to a general DO-covariant quantum channel. Interestingly, a simple analysis shows that if a DU-covariant superchannel $\Delta$ maps bit-flip channels to bit-flip channels, then all bit-flip channels are fixed points of $\Delta$.} 

\end{Example}

\begin{Example}[Pauli channels]

{\em 
The Pauli channel is defined via
\[
\Phi(\rho)=  \sum_{\alpha=0}^3 p_\alpha \sigma_\alpha\rho \sigma_\alpha ,
\]
with $\sigma_0 =\oper_2$, and $p_\alpha$ stands for a probability vector. It belongs to a class of diagonal orthogonal maps \cite{Bihalan}. The corresponding Choi matrix reads

\[
C_\Phi=
\begin{pmatrix}
p_0+p_3 & 0 & 0 & p_0-p_3\\[4pt]
0 & p_1+p_2 & p_1-p_2 & 0\\[4pt]
0 & p_1-p_2 & p_1+p_2 & 0\\[4pt]
p_0-p_3 & 0 & 0 & p_0+p_3
\end{pmatrix} ,
\]
and hence,

$$  \Delta_\Theta(C_\Phi) =
\begin{pmatrix}
p_{11} & 0 & 0 & D_{11,22}\,(p_0-p_3)\\[4pt]
0 & p_{12} & D_{12,21}\,(p_1-p_2) & 0\\[4pt]
0 & D_{21,12}\, (p_1-p_2) & p_{21} & 0\\[4pt]
D_{22,11}\,(p_0-p_3) & 0 & 0 & p_{22}
\end{pmatrix} ,
$$
where

\begin{eqnarray}
    p_{11} &=& (A_{11,11}+A_{11,22})(p_{0}+p_{3})+(A_{11,12}+A_{11,21})(p_{1}+p_{2}),  \nonumber \\
    p_{12} &=& (A_{12,11}+A_{12,22})(p_{0}+p_{3})+(A_{12,12}+A_{12,21})(p_{1}+p_{2}),\\
    p_{21} &=& (A_{21,11}+A_{21,22})(p_{0}+p_{3})+(A_{21,12}+A_{21,21})(p_{1}+p_{2}), \nonumber\\
    p_{22}&=&(A_{22,11}+A_{22,22})(p_{0}+p_{3})+(A_{22,12}+A_{22,21})(p_{1}+p_{2}). \nonumber
\end{eqnarray}
}
{This shows that any DU-covariant superchannel maps a Pauli channel to a DO-covariant channel. In the next section we introduce a class of DO-covariant superchannels which map Pauli channels to Pauli channels. }
\end{Example}

{

Finally, observe that dephasing superchannels with 1-parameter

\begin{equation}
    \mathbb{M} (\lambda)  =  \begin{pmatrix}
1 & \lambda & \lambda & 1\\[4pt]
\lambda & 1 & 1 & \lambda\\[4pt]
\lambda & 1 & 1 & \lambda\\[4pt]
 1 & \lambda &\lambda & 1
\end{pmatrix} \ , \ \ \ \ |\lambda|\leq 1 ,
\end{equation}
keep all DO-covariant qubit channels invariant (DO-covariant qubit channels have $X$-shape Choi matrices).

}

\section{Pauli Superchannels}\label{sec9}

In analogy to Pauli quantum channels, let us define a Pauli superchannel via its representing map

\begin{equation}   \label{Super-P}
    \Delta(X)=\sum_{\mu,\nu=0}^3 \pi_{\mu\nu}\,
\sigma_\mu\otimes\sigma_\nu\,X\,\sigma_\mu\otimes\sigma_\nu ,
\end{equation}
where $\pi_{\mu\nu} \geq 0$, and $\sum_{\mu,\nu} \pi_{\mu\nu}= 1$. This gives the Pauli superchannel as follows:

\begin{equation}
    \Theta[\Phi] = \sum_{\mu,\nu=0}^3 \pi_{\mu\nu}\, \Sigma_\nu \circ \Phi \circ \Sigma_\mu , 
\end{equation}
where   $\Sigma_\mu(X) = \sigma_\mu X \sigma_\mu$. 
 One easily finds that

\begin{equation}
    {\rm Tr}_{B_1} \Delta(X) = \widetilde{\Delta}( {\rm Tr}_{A_1} X) ,
\end{equation}
with  $\widetilde{\Delta}(A) = \sum_\mu \pi_\mu \sigma_\mu A \sigma_\mu$, where $\pi_\mu = \sum_\nu \pi_{\mu\nu}$. 

\begin{proposition} A Pauli superchannel is DO-covariant. Moreover, it is diagonal unitary covariant if and only if $\pi_{\mu\nu}$ satisfy the additional constraints

\begin{equation}   \label{pipi}
    \pi_{\alpha 1} = \pi_{\alpha 2} \ , \quad \pi_{1 \alpha } = \pi_{2 \alpha }
\end{equation}
for $\alpha=0,1,2,3$. 
    
\end{proposition}
{

\begin{proof}
Note that if $O$ is a $2 \times 2$ diagonal orthogonal matrix, i.e. $O={\rm diag}[o_1,o_2]$ with $o_i = \pm 1$, then 

\[  O \sigma_\mu O = s_\mu(O) \sigma_\mu ,\]
where

\[  s_0(O) = s_3(O) = 1 \ , \ \ \ s_1(O) = s_2(O) = o_1 o_2 .  \]
Consequently, we have

\begin{eqnarray*}
 &&    O_1 \otimes O_2 \Delta(O_1 \otimes O_2 X O_1 \otimes O_2) O_1 \otimes O_2 =  \sum_{\mu,\nu=0}^3 \pi_{\mu\nu}\,
O_1 \sigma_\mu O_1 \otimes O_2 \sigma_\nu O_2 \,X\,O_1 \sigma_\mu O_1 \otimes O_2 \sigma_\nu O_2 \\
&& = \sum_{\mu,\nu=0}^3 \pi_{\mu\nu}\, |s_\mu(O)|^2 |s_\nu(O)|^2 \,
\sigma_\mu\otimes\sigma_\nu\,X\,\sigma_\mu\otimes\sigma_\nu = \Delta(X).
\end{eqnarray*}
The last equality follows from the fact that $|s_\mu(O)|^2 = |s_\nu(O)|^2 =1$. This proves that a Pauli superchannel is DO-covariant.

Now we impose diagonal unitary covariance. For qubits, the DU weight spaces are
\[
V_{00},\quad V_{0;12},\quad V_{0;21},\quad
V_{12;0},\quad V_{21;0},\quad
V_{12;12},\quad V_{12;21},\quad V_{21;12},\quad V_{21;21}.
\]
Thus, $\Delta$ is DU-covariant iff each of these spaces is invariant.
Using
\[
\sigma_0e_{12}\sigma_0=e_{12},\qquad
\sigma_1e_{12}\sigma_1=e_{21},\qquad
\sigma_2e_{12}\sigma_2=-e_{21},\qquad
\sigma_3e_{12}\sigma_3=-e_{12},
\]
and the analogous relations for $e_{21}$, one easily computes $\Delta(e_{ij} \otimes e_{ab})$. In particular, we obtain

\[
\begin{aligned}
\Delta(e_{11}\otimes e_{12})
& =
(\pi_{00}-\pi_{03}+\pi_{30}-\pi_{33})
\,e_{11}\otimes e_{12} +
(\pi_{01}-\pi_{02}+\pi_{31}-\pi_{32})
\,e_{11}\otimes e_{21}
\\
& + 
(\pi_{10}-\pi_{13}+\pi_{20}-\pi_{23})
\,e_{22}\otimes e_{12}
+
(\pi_{11}-\pi_{12}+\pi_{21}-\pi_{22})
\,e_{22}\otimes e_{21} \ ,
\end{aligned}
\]
and hence, invariance of $V_{0;12}$ gives the following constraints:

\begin{equation}
\pi_{01}-\pi_{02}+\pi_{31}-\pi_{32}=0,
\qquad
\pi_{11}-\pi_{12}+\pi_{21}-\pi_{22}=0.
\end{equation}
Consider now the space $V_{12;12}$. Here we have
\[
\begin{aligned}
\Delta(e_{12}\otimes e_{12})
& = 
(\pi_{00}-\pi_{03}-\pi_{30}+\pi_{33})
\,e_{12}\otimes e_{12}
+
(\pi_{10}-\pi_{13}-\pi_{20}+\pi_{23})
\,e_{21}\otimes e_{12}
\\
& +
(\pi_{01}-\pi_{02}-\pi_{31}+\pi_{32})
\,e_{12}\otimes e_{21}
+
(\pi_{11}-\pi_{12}-\pi_{21}+\pi_{22})
\,e_{21}\otimes e_{21}.
\end{aligned}
\]
Hence invariance of the one-dimensional space $V_{12;12}$ gives rise to additional constraints
\begin{equation}
\pi_{01}-\pi_{02}-\pi_{31}+\pi_{32}=0,
\qquad
\pi_{11}-\pi_{12}-\pi_{21}+\pi_{22}=0.
\end{equation}
Adding and subtracting these equations yields
\begin{equation}
\pi_{\alpha 1}=\pi_{\alpha 2},
\qquad
\alpha=0,1,2,3.
\end{equation}
Similarly, invariance of $V_{12;0}$ implies
\begin{equation}
\pi_{10}-\pi_{20}+\pi_{13}-\pi_{23}=0,
\qquad
\pi_{11}-\pi_{21}+\pi_{12}-\pi_{22}=0,
\end{equation}
whereas invariance of $V_{21;21}$ gives additionally
\begin{equation}
\pi_{10}-\pi_{20}-\pi_{13}+\pi_{23}=0,
\qquad
\pi_{11}-\pi_{21}-\pi_{12}+\pi_{22}=0.
\end{equation}
Hence
\begin{equation}
\pi_{1\alpha}=\pi_{2\alpha},
\qquad
\alpha=0,1,2,3.
\end{equation}
Conversely, if these relations hold, then it is clear that $\Delta$ is DU-covariant.  
\end{proof}

\begin{remark}
    In Appendix \ref{APP-Pauli} we provide the structure of $\{A,B,C,D,E,P,Q,R,S\}$ matrices in terms of $\pi_{\mu\nu}$. A detailed analysis of these matrices shows that $E=P=Q=R=S=0$ if and only if the conditions (\ref{pipi}) hold.  
\end{remark}

}
Note that, the constraints (\ref{pipi}) provide generalization of the well known constraint for Pauli channels. A Pauli channel

\begin{equation}
    \Phi(X) = \sum_{\mu=0}^3 p_\mu \, \sigma_\mu X \sigma_\mu ,
\end{equation}
is diagonal orthogonal covariant, and it is diagonal unitary covariant only if, additionally,  $p_1=p_2$ \cite{Bihalan}. 
Hence, for diagonal unitary covariance, the $\pi_{\mu\nu}$ matrix {is required to have} the following structure

\[ \pi_{\mu\nu} = 
\begin{pNiceMatrix}[margin, extra-margin=2pt] 
\pi_{00} & \pi_{01} & \pi_{02} & \pi_{03} \\
\pi_{10} & \pi_{11} & \pi_{12} & \pi_{13} \\
\pi_{20} & \pi_{21} & \pi_{22} & \pi_{23} \\
\pi_{30} & \pi_{31} & \pi_{32} & \pi_{33}
\CodeAfter
  \begin{tikzpicture}[black, thick]
  
    \draw ([xshift=-2pt, yshift=2pt]1-2.north west) rectangle ([xshift=2pt, yshift=-2pt]1-3.south east);

    \draw ([xshift=-2pt, yshift=2pt]2-1.north west) rectangle ([xshift=2pt, yshift=-2pt]3-1.south east);

    \draw ([xshift=-2pt, yshift=2pt]2-2.north west) rectangle ([xshift=2pt, yshift=-2pt]3-3.south east);

    \draw ([xshift=-2pt, yshift=2pt]2-4.north west) rectangle ([xshift=2pt, yshift=-2pt]3-4.south east);

    \draw ([xshift=-2pt, yshift=2pt]4-2.north west) rectangle ([xshift=2pt, yshift=-2pt]4-3.south east);
    
  \end{tikzpicture}
\end{pNiceMatrix} ,
\]
where elements within the same block are equal. 
 Now, consider the action of $\Delta$ on the Choi matrix of a Pauli channel

$$ C_\Phi = \sum_{\alpha=0}^3 p_\alpha |B_\alpha\>\<B_\alpha| ,  $$
where $|B_\alpha\> = (\oper_2 \otimes \sigma_\alpha) \sum_{i=1}^2 |e_i \otimes e_i\>$ are (unnormalized) Bell states. It can be shown that

\begin{equation}
    \Delta(C_\Phi) = \sum_{\alpha=0}^3 q_\alpha |B_\alpha\>\<B_\alpha| ,
\end{equation}
where

\begin{equation} \label{qMp}
    q_\alpha = \sum_{\beta=0}^3 M_{\alpha\beta}\, p_\beta ,
\end{equation}
and

\begin{equation}  \label{M}
    M_{\alpha\beta} = \sum_{\alpha,\beta; \, \alpha\oplus\beta = \mu\oplus\nu} \pi_{\mu\nu} ,
\end{equation}
where $\alpha\oplus\beta$ denotes 2-bit XOR operation:

\[
\begin{array}{c|cccc}
\alpha \oplus \beta & 0 & 1 & 2 & 3 \\ \hline
0 & 0 & 1 & 2 & 3 \\
1 & 1 & 0 & 3 & 2 \\
2 & 2 & 3 & 0 & 1 \\
3 & 3 & 2 & 1 & 0
\end{array} .
\]
Hence, a Pauli superchannel (\ref{Super-P}) transforms a Pauli channel parameterized by $p_\alpha$ into a Pauli channel parameterized by $q_\alpha$ via (\ref{qMp}), with $M_{\alpha\beta}$ being a bistochastic matrix defined in (\ref{M}).

\section{Conclusions and outlook}\label{sec10} 

In this work, we have developed a comprehensive framework for the characterization of diagonal unitary covariant (DU-covariant) quantum superchannels, extending symmetry-based methods from quantum channels to higher-order quantum transformations. Starting from the covariance conditions at the level of Choi operators, we derived a complete parametrization of all DU-covariant superchannels and identified the compatibility constraints required for complete positivity and trace
preservation. It is shown that any DU-covariant superchannel admits a decomposition into four canonical components, each acting on different diagonal and off-diagonal blocks of the underlying operator space. This structural decomposition provides a transparent way to understand how symmetry restricts the action of superchannels and enables explicit computations in concrete scenarios.  As a consequence, the action of such superchannels admits a transparent interpretation in terms of classical stochastic processing of diagonal components combined with controlled transformations of coherence sectors. This block structure generalizes known results for DU-covariant quantum channels \cite{Singh_2021}  and highlights the intricate relation between channel-level and superchannel-level covariance.

A set of DU-covariant superchannels is closed under the composition of supermaps. We have provided explicit composition rules for their canonical components. This result is important from the operational perspective as it  enables systematic constructions of complex transformations from elementary building blocks. In particular, the composition rules clarify how the classical stochastic parts defined by $\Delta_1$ and coherence-related contributions $\{\Delta_2,\Delta_2,\Delta_3\}$ combine. In Section 6 we provided a similar analysis for diagonal orthogonally covariant (DO) superchannels.  Here, the analysis is more demanding since any such supermap can be decomposed into nine canonical components with intricate composition rules. Table~2 provides the number of real parameters characterizing DU and DO-covariant supermaps before and after TP preserving conditions. For comparison, we also included the number of parameters characterizing DU and DO-covariant maps.

\begin{table}[t]
\centering
\small
\label{tab:parameter_count}
\begin{tabular}{|>{\centering\arraybackslash}p{3cm}|>{\centering\arraybackslash}p{4cm}|>{\centering\arraybackslash}p{4.3cm}|}
\hline
& & \\
\textbf{{Class of (super)maps}} & \textbf{{Number of parameters for HP}} & \textbf{{Number of parameters for HP+TP}} \\
& & \\
\hline
& & \\
{DU-covariant maps}
& {$n_{\rm DU} = d(2d-1)$}
& {$n_{\rm DU}^{\rm TP} = 2d(d-1)$} \\
& & \\
\hline
& & \\
{DO-covariant maps}
& {$n_{\rm DO} = d(3d-2)$}
& {$n_{\rm DO}^{\rm TP} = 3d(d-1)$} \\
& & \\
\hline
& & \\
{DU-covariant supermaps}
& {$N_{\rm DU} = n_{\rm DU} \, n_{\rm DU}$}
& {$N_{\rm DU}^{\rm TP} = n_{\rm DU}^{\rm TP} ( n_{\rm DU} + 1)$} \\
& & \\
\hline
& & \\
{DO-covariant supermaps}
& {$N_{\rm DO} = n_{\rm DO}\, n_{\rm DO}$}
& {$N_{\rm DO}^{\rm TP} = n_{\rm DO}^{\rm TP} ( n_{\rm DO} + 1)$} \\
& & \\
\hline
\end{tabular}
\caption{{The number of independent real parameters for DU- and DO-covariant (super)maps. The second column characterizes Hermiticity-preserving (HP) (super)maps, and the third column gives the number of free parameters after imposing the trace-preserving (TP) conditions.}}
\end{table}

An important application of the general theory developed here is the analysis of dephasing superchannels, which define a natural subclass of DU-covariant superchannels. By embedding dephasing superchannels into the DU-covariant framework, we have shown that their defining properties -- suppression of coherence transfer and preservation of diagonal structure -- emerge directly from diagonal unitary symmetry. This unifies recent results on dephasing superchannels \cite{Puchala2021,Korzekwa2018} with a broader covariance-based perspective developed in the present work. 

The examples presented for qubit channels further illustrate the practical relevance of our results. In particular, we have demonstrated how DU-covariant superchannels act on well studied  noise models such as amplitude-damping, bit-flip, and Pauli channels. These examples show explicitly how symmetry constraints restrict admissible transformations and how classical stochastic matrices governing diagonal blocks determine the effective noise reshaping at the channel level. Such explicit constructions underline the usefulness of the framework for concrete applications in noise manipulation and channel engineering.

Beyond the structural characterization, these results provide a practical toolkit for studying symmetry-restricted transformations of quantum processes. Since diagonal unitary covariance appears in a wide range of contexts, from decoherence models to phase-covariant dynamics and classical-quantum hybrid processes, our framework offers a versatile starting point for future investigations. The methods developed in this paper might be, in principle, extended to block-diagonal unitary (and orthogonal) matrices potentially leading to new classifications of physically relevant superchannels. The explicit block structure of DU-covariant superchannels provides a natural platform for studying resource theories of processes. For example, to study coherence  at the level of channels rather than states.  An interesting direction for future work is to use the present characterization of DU-covariant superchannels to probe the celebrated PPT$^2$ conjecture \cite{Christandl2012, Christandl2019}. Recently, it was shown \cite{Singh2022} that the PPT$^2$ conjecture holds for diagonal unitary covariant quantum channels. It would be very interesting to prove it also for DU-covariant superchannels. 

{An interesting direction for future work is to extend the present analysis to general process tensors (quantum combs) \cite{Chiribella,Pollock2018, Milz2021Quantum, Pollock2018NonMarkovian,Milz2024Characterising, Jencova2024Structure, Jencova2026Order, Zambon2024Process}. Since a superchannel is precisely a one-slot quantum comb, the 
representation-theoretic approach developed here naturally generalizes to multi-time processes. The Choi operator of an $N$-slot process tensor carries a tensor-product representation of the symmetry group acting independently on each input and output time, and covariance can again be formulated as an intertwining condition. If \(R^{(N)}\) denotes the Choi operator of an \(N\)-slot process tensor the covariance can be imposed by requiring
\[
\left[
R^{(N)},
\,U_{0}\otimes \cdots \otimes U_N \otimes \overline{U}_{0}\otimes\cdots\otimes
\overline{U}_{N}
\right]
=0,
\]
for all (diagonal) unitaries \(U_{k}\). $N=0$ corresponds to covariant quantum channels, whereas $N=1$ corresponds to covariant quantum superchannels.  One therefore expects a canonical block decomposition of covariant process tensors analogous to the one obtained here for superchannels, with complete positivity reducing to positivity of the corresponding invariant blocks and the causality constraints becoming linear relations between them. Such a classification would provide a systematic representation-theoretic framework for symmetry-restricted quantum combs and process tensors.
We leave the detailed analysis for future research. }

\section*{Acknowledgement}

D.C. was supported by the Polish National Science
Center project No. 2024/55/B/ST2/01781. V.P. and Sohail thank Abhay Srivastav for discussions. D.C thanks Tanmay Singal for the discussion at the early stage of this project.  

\subsection*{ Conflicts of interests/Competing interests}
 The authors declare that they have no conflict of interest.

\appendix

\section*{Appendix}

\numberwithin{equation}{section}

%%%%%%%%%%%%%%%%%%%%%%%%%%%%%%%%%%%%%%%%%%%%%%%%%%%%%%%%%%%%%
\section{Proof of Proposition \ref{PRO-UV}}\label{proof_prop3}

Define the maximally entangled vector in $\HAa \otimes \HAa$
\begin{equation*}
|\psi^+\rangle = \sum_i e_i \otimes e_i.
\end{equation*}
Then the Choi of $\Phi$ reads
\begin{equation*}
C_\Phi = (\mathrm{id} \otimes \Phi)(|\psi^+\rangle \langle\psi^+|) .
\end{equation*}
Let us define 

\begin{equation}
{\Phi}_g(X)
:= V^\dagger_g \, \Phi(U_g (X) U^\dagger_g)\, V_g .
\end{equation}
By covariance, $\Phi_g = \Phi$ and hence $C_{\Phi_g} = C_\Phi$. Now, let us compute the Choi matrix of $\Phi_g$:

\begin{align*}
C_{\Phi_g}
&= (\mathrm{id} \otimes \Phi_g)(|\psi^+\rangle\langle\psi^+|) \\
&= (\oper \otimes V^\dagger_g)
   (\mathrm{id} \otimes \Phi)
   \big[(\oper \otimes U_g)\,|\psi^+\rangle\langle\psi^+|\,
        (\oper \otimes U^\dagger_g)\big]
   (\oper \otimes V_g).
\end{align*}
For any operator $X$ on $\HAa \otimes\HAa$ 

\begin{equation}  \label{XXT}
(X \otimes \oper)|\psi^+\rangle = (\oper \otimes X^T)|\psi^+\rangle ,
\end{equation}
and hence

\begin{equation*}
(\oper \otimes U_g)|\psi^+\rangle
= ({U}^T_g \otimes \oper)|\psi^+\rangle.
\end{equation*}
Thus
\begin{align}
(\oper \otimes U_g)\,|\psi^+\rangle\langle\psi^+|\,
(\oper \otimes U^\dagger_g)
&= ({U}^T_g \otimes \oper)\,
   |\psi^+\rangle\langle\psi^+|\,
   (\overline{U}_g \otimes \oper).
\end{align}
Applying $(\mathrm{id} \otimes \Phi)$ and then
$(\oper \otimes V^\dagger_g)$, we get
\begin{align}
C_{\Phi_g}
&= (U_g^T \otimes V^\dagger_g)\, C_\Phi \, (\overline{U}_g \otimes V_g) ,
\end{align}
or, equivalently 

$$  \overline{U}_g \otimes V_g \, C_\Phi \, (\overline{U}_g \otimes V_g)^\dagger = C_\Phi , $$
due to $C_{\Phi_g} = C_\Phi$.

\section{Proof of Proposition \ref{PRO-MN}} \label{PROOF-MN}

The formula (\ref{XXT}) may be generalized as follows: fixing orthonormal bases $\{e_i\}$ and $\{f_a\}$ in $\HAa$ and $\HAb$, respectively, one has for an arbitrary operator $X : \HAa \to \HAb$

\begin{equation}
    (\oper_{A_0} \otimes X)|\psi^+\>_{A_0} = (X^T \otimes \oper_{A_1}) |\psi^+\>_{A_1} ,
\end{equation}
where $|\psi^+\>_{A_0} = \sum_i e_i \otimes e_i$ and $|\psi^+\>_{A_1} = \sum_a f_a \otimes f_a$, and the transposition is defined w.r.t. $\{e_i\}$ and $\{f_a\}$, i.e.

$$  \<e_i|X^T|f_a\> := \<f_a|X|e_i\> = X_{ai} .  $$
Hence one obtains the following formula  for the Choi matrix of a map $\Phi : \BAa \to \BAb$

\begin{equation}
    C_\Phi = ({\rm id}_{A_0} \otimes \Phi)(|\psi^+\>_{A_0}\<\psi^+|) = (\mathbb{T}[{{\Phi^\ddag} }] \otimes {\rm id}_{A_1} )(|\psi^+\>_{A_1}\<\psi^+|) .
\end{equation}
Indeed, any Hermiticity preserving $\Phi$ can be represented as $\Phi(X) = \sum_\alpha \lambda_\alpha K_\alpha X K_\alpha^\dagger$ with real $\lambda_\alpha$. Hence

\begin{eqnarray*}
    C_\Phi &=& ({\rm id}_{A_0} \otimes \Phi)(|\psi^+\>_{A_0}\<\psi^+|) = \sum_\alpha \lambda_\alpha \, (\oper_{A_0} \otimes K_\alpha)|\psi^+\>_{A_0}\<\psi^+|(\oper_{A_0} \otimes K_\alpha^\dagger) \\
    &=& \sum_\alpha \lambda_\alpha \, (K_\alpha^T \otimes \oper_{A_1})|\psi^+\>_{A_1}\<\psi^+| (K_\alpha^T \otimes\oper_{A_1})^\dagger = (\mathbb{T}[{\Phi^\ddag }] \otimes {\rm id}_{A_1} )(|\psi^+\>_{A_1}\<\psi^+|) 
    %(\mathbb{T}[{\Phi^\ddager}] \otimes {\rm id}_{A_1} )(|\psi^+\>_{A_1}\<\psi^+|) .
\end{eqnarray*}
due to

$$ {\Phi^\ddag}(X) = \sum_\alpha \lambda_\alpha \, K_\alpha^\dagger X {K}_\alpha \ , \quad   \mathbb{T}[\Phi^\ddag](X) = \sum_\alpha \lambda_\alpha \, K_\alpha^T X \overline{K}_\alpha . $$
Now, $\Theta[\Phi] = \mathcal{N}_1 \circ \Phi \circ \mathcal{N}^{\ddag}_{0}$ can be represented as
$\Theta = \Theta_1 \circ \Theta_0 = \Theta_0 \circ \Theta_1 $ with

$$  \Theta_1[\Phi] = \mathcal{N}_1 \circ \Phi \ , \quad \Theta_0[\Phi] = \Phi \circ \mathcal{N}^{\ddag}_{0}  \ .$$
Hence, $\Delta_\Theta = \Delta_1 \circ \Delta_0 = \Delta_0 \circ \Delta_1$ and $\Delta_i$ is a  representing map of $\Theta_i$. One has

\begin{eqnarray*}
    \Delta_1(C_\Phi) &=& C_{\Theta_1[\Phi]} = ({\rm id}_{A_0} \otimes \mathcal{N}_1 \circ \Phi)(|\psi^+\>_{A_0}\<\psi^+|) =  ({\rm id}_{A_0} \otimes \mathcal{N}_1)({\rm id}_{A_0} \otimes  \Phi)(|\psi^+\>_{A_0}\<\psi^+|) \\
    &=& ({\rm id}_{A_0} \otimes \mathcal{N}_1)(C_\Phi) .
\end{eqnarray*}
Similarly, 

\begin{eqnarray*}
    \Delta_0(C_\Phi) &=& C_{\Theta_1[\Phi]} = ({\rm id}_{B_0} \otimes \Phi\circ \mathcal{N}_0^\ddag)(|\psi^+\>_{B_0}\<\psi^+|) =  (\mathbb{T}[\mathcal{N}_0 \circ {\Phi^\ddag}] \otimes {\rm id}_{A_1})(|\psi^+\>_{A_1}\<\psi^+|) \\
    &=&  (\mathbb{T}[\mathcal{N}_0] \otimes {\rm id}_{A_1})  (\mathbb{T}[{\Phi^\ddag}] \otimes {\rm id}_{A_1})(|\psi^+\>_{A_1}\<\psi^+|) = (\mathbb{T}[\mathcal{N}_0] \otimes {\rm id}_{A_1})(C_\Phi) ,
\end{eqnarray*}
where we have used

$$  (\Phi \circ \Psi)^\ddag = \Psi^\ddag \circ {\Phi^\ddag}\ , \quad \mathbb{T}[\Phi \circ \Psi] =   \mathbb{T}[\Phi] \circ  \mathbb{T}[\Psi]\ . $$
%Joni Mitchell with James Taylor - You Can Close Your Eyes
Hence,

\begin{equation}
    \Delta_1 = {\rm id}_{A_0} \otimes \mathcal{N}_1\ , \quad \Delta_0 = \mathbb{T}[\mathcal{N}_0] \otimes {\rm id}_{A_1} .
\end{equation}

\section{Covariance of Pauli superchannels} \label{APP-Pauli}

{

Define
\[
\begin{aligned}
a_0&=\pi_{00}+\pi_{03}+\pi_{30}+\pi_{33},&
a_1&=\pi_{01}+\pi_{02}+\pi_{31}+\pi_{32},\\
a_2&=\pi_{10}+\pi_{13}+\pi_{20}+\pi_{23},&
a_3&=\pi_{11}+\pi_{12}+\pi_{21}+\pi_{22},
\end{aligned}
\]
and
\[
\begin{aligned}
b_0&=\pi_{00}-\pi_{03}+\pi_{30}-\pi_{33},&
b_1&=\pi_{10}-\pi_{13}+\pi_{20}-\pi_{23},\\
c_0&=\pi_{00}+\pi_{03}-\pi_{30}-\pi_{33},&
c_1&=\pi_{01}+\pi_{02}-\pi_{31}-\pi_{32},\\
e_0&=\pi_{10}+\pi_{13}-\pi_{20}-\pi_{23},&
e_1&=\pi_{11}+\pi_{12}-\pi_{21}-\pi_{22}, \\
r_0&=\pi_{01}-\pi_{02}+\pi_{31}-\pi_{32},&
r_1&=\pi_{11}-\pi_{12}+\pi_{21}-\pi_{22},
\end{aligned}
\]
and 

\begin{eqnarray*}
 d&=\pi_{00}-\pi_{03}-\pi_{30}+\pi_{33}, \\
 p&=\pi_{10}-\pi_{13}-\pi_{20}+\pi_{23},\\
q&=\pi_{11}-\pi_{12}-\pi_{21}+\pi_{22}, \\
s&=\pi_{01}-\pi_{02}-\pi_{31}+\pi_{32}.
\end{eqnarray*}
Then the matrices corresponding to the Pauli superchannel have the following form:

\begin{itemize}
    \item $A$-sector

\[
A=
\begin{pmatrix}
a_0&a_1&a_2&a_3\\
a_1&a_0&a_3&a_2\\
a_2&a_3&a_0&a_1\\
a_3&a_2&a_1&a_0
\end{pmatrix},
\]

\item $(B,R)$-sector

\[
B=
\begin{pmatrix}
0&b_0&0&b_1\\
b_0&0&b_1&0\\
0&b_1&0&b_0\\
b_1&0&b_0&0
\end{pmatrix},
\qquad
R=
\begin{pmatrix}
0&r_0&0&r_1\\
r_0&0&r_1&0\\
0&r_1&0&r_0\\
r_1&0&r_0&0
\end{pmatrix},
\]

\item $(C,E)$-sector
\[
C=
\begin{pmatrix}
0&0&c_0&c_1\\
0&0&c_1&c_0\\
c_0&c_1&0&0\\
c_1&c_0&0&0
\end{pmatrix}, \quad
E=
\begin{pmatrix}
0&0&e_0&e_1\\
0&0&e_1&e_0\\
e_0&e_1&0&0\\
e_1&e_0&0&0
\end{pmatrix},
\]

\item $(D,P,Q,S)$-sector
\[
D=
\begin{pmatrix}
0&0&0&d\\
0&0&d&0\\
0&d&0&0\\
d&0&0&0
\end{pmatrix},
\
P=
\begin{pmatrix}
0&0&0&p\\
0&0&p&0\\
0&p&0&0\\
p&0&0&0
\end{pmatrix},
\
Q=
\begin{pmatrix}
0&0&0&q\\
0&0&q&0\\
0&q&0&0\\
q&0&0&0
\end{pmatrix},
\
S=
\begin{pmatrix}
0&0&0&s\\
0&0&s&0\\
0&s&0&0\\
s&0&0&0
\end{pmatrix}.
\]

\end{itemize}
It is therefore clear that $E=R=P=Q=S=0$ if only if conditions (\ref{pipi}) hold. This analysis clearly shows that (\ref{pipi}) implies that at each sector we have only one nontrivial matrix. 

DO-covariant qubit channel is represented via
\[
\Phi(X)
=
\sum_{i,j=1}^{2}
a_{ij}\,e_{ij}Xe_{ji}
+
\sum_{{i\neq j}}
b_{ij}\,e_{ii}Xe_{jj}
+
\sum_{{i\neq j}}
c_{ij}\,e_{ii}X^{T}e_{jj},
\]
and hence for a Pauli channel $\Phi(X) = \sum_\mu p_\mu \sigma_\mu X \sigma_\mu$ the  corresponding matrices are defined as follows
\[
a=
\begin{pmatrix}
p_{0}+p_{3} & p_{1}+p_{2}\\
p_{1}+p_{2} & p_{0}+p_{3}
\end{pmatrix},
\]
together with
\[
b=
\begin{pmatrix}
0 & p_{0}-p_{3}\\
p_{0}-p_{3} & 0
\end{pmatrix},
\qquad
c=
\begin{pmatrix}
0 & p_{1}-p_{2}\\
p_{1}-p_{2} & 0
\end{pmatrix}.
\]
Therefore, the Pauli channel is DU-covariant, i.e. $c_{ij}=0$,  if and only if $p_1=p_2$.

}

\bibliographystyle{unsrturl}
\bibliography{name}{}

\begin{thebibliography}{10}

\bibitem{QIT}
Michael~A. Nielsen and Isaac~L. Chuang.
\newblock {\em Quantum Computation and Quantum Information}.
\newblock Cambridge University Press, June 2012.
\newblock \href {https://doi.org/10.1017/cbo9780511976667} {\path{doi:10.1017/cbo9780511976667}}.

\bibitem{Wilde2013}
Mark~M. Wilde.
\newblock {\em Quantum Information Theory}.
\newblock Cambridge University Press, Cambridge, May 2013.
\newblock \href {https://doi.org/10.1017/CBO9781139525343} {\path{doi:10.1017/CBO9781139525343}}.

\bibitem{Werner1989}
Reinhard~F. Werner.
\newblock Quantum states with {E}instein-{P}odolsky-{R}osen correlations admitting a hidden-variable model.
\newblock {\em Physical Review A}, 40:4277--4281, Oct 1989.
\newblock URL: \url{https://link.aps.org/doi/10.1103/PhysRevA.40.4277}, \href {https://doi.org/10.1103/PhysRevA.40.4277} {\path{doi:10.1103/PhysRevA.40.4277}}.

\bibitem{Horodecki1999}
Micha\l{} Horodecki and Pawe\l{} Horodecki.
\newblock Reduction criterion of separability and limits for a class of distillation protocols.
\newblock {\em Physical Review A}, 59:4206--4216, Jun 1999.
\newblock URL: \url{https://link.aps.org/doi/10.1103/PhysRevA.59.4206}, \href {https://doi.org/10.1103/PhysRevA.59.4206} {\path{doi:10.1103/PhysRevA.59.4206}}.

\bibitem{HHHH}
Ryszard Horodecki, Pawe\l{} Horodecki, Micha\l{} Horodecki, and Karol Horodecki.
\newblock Quantum entanglement.
\newblock {\em Reviews of Modern Physics}, 81:865--942, Jun 2009.
\newblock URL: \url{https://link.aps.org/doi/10.1103/RevModPhys.81.865}, \href {https://doi.org/10.1103/RevModPhys.81.865} {\path{doi:10.1103/RevModPhys.81.865}}.

\bibitem{VollbrechtWerner2001}
Karl G.~H. Vollbrecht and Reinhard~F. Werner.
\newblock Entanglement measures under symmetry.
\newblock {\em Physical Review A}, 64(6):062307, 2001.
\newblock \href {https://doi.org/10.1103/PhysRevA.64.062307} {\path{doi:10.1103/PhysRevA.64.062307}}.

\bibitem{Chruscinski2006a}
Dariusz Chru\'sci\'nski and Andrzej Kossakowski.
\newblock Multipartite invariant states. {I}. unitary symmetry.
\newblock {\em Physical Review A}, 73:062314, Jun 2006.
\newblock URL: \url{https://link.aps.org/doi/10.1103/PhysRevA.73.062314}, \href {https://doi.org/10.1103/PhysRevA.73.062314} {\path{doi:10.1103/PhysRevA.73.062314}}.

\bibitem{Chruscinski2006b}
Dariusz Chru\'sci\'nski and Andrzej Kossakowski.
\newblock Multipartite invariant states. {II}. orthogonal symmetry.
\newblock {\em Physical Review A}, 73:062315, Jun 2006.
\newblock URL: \url{https://link.aps.org/doi/10.1103/PhysRevA.73.062315}, \href {https://doi.org/10.1103/PhysRevA.73.062315} {\path{doi:10.1103/PhysRevA.73.062315}}.

\bibitem{Scutaru1979}
Horia Scutaru.
\newblock Some remarks on covariant completely positive linear maps on {$C^*$}-algebras.
\newblock {\em Reports on Mathematical Physics}, 16(1):79--84, August 1979.
\newblock \href {https://doi.org/10.1016/0034-4877(79)90040-5} {\path{doi:10.1016/0034-4877(79)90040-5}}.

\bibitem{Holevo1993}
Alexander~S. Holevo.
\newblock A note on covariant dynamical semigroups.
\newblock {\em Reports on Mathematical Physics}, 32:211, 1993.
\newblock \href {https://doi.org/10.1016/0034-4877(93)90014-6} {\path{doi:10.1016/0034-4877(93)90014-6}}.

\bibitem{Holevo1996}
Alexander~S. Holevo.
\newblock Covariant quantum markovian evolutions.
\newblock {\em Journal of Mathematical Physics}, 37:1812, 1996.
\newblock \href {https://doi.org/10.1063/1.531481} {\path{doi:10.1063/1.531481}}.

\bibitem{Singh_2021}
Satvik Singh and Ion Nechita.
\newblock Diagonal unitary and orthogonal symmetries in quantum theory.
\newblock {\em Quantum}, 5:519, August 2021.
\newblock URL: \url{http://dx.doi.org/10.22331/q-2021-08-09-519}, \href {https://doi.org/10.22331/q-2021-08-09-519} {\path{doi:10.22331/q-2021-08-09-519}}.

\bibitem{Nechita_2021}
Ion Nechita and Satvik Singh.
\newblock A graphical calculus for integration over random diagonal unitary matrices.
\newblock {\em Linear Algebra and its Applications}, 613:46–86, March 2021.
\newblock URL: \url{http://dx.doi.org/10.1016/j.laa.2020.12.014}, \href {https://doi.org/10.1016/j.laa.2020.12.014} {\path{doi:10.1016/j.laa.2020.12.014}}.

\bibitem{Singh2022}
Satvik Singh and Ion Nechita.
\newblock The {PPT} square conjecture holds for all choi-type maps.
\newblock {\em Annales Henri Poincar{\'e}}, 23(9), March 2022.
\newblock URL: \url{http://dx.doi.org/10.1007/s00023-022-01166-0}, \href {https://doi.org/10.1007/s00023-022-01166-0} {\path{doi:10.1007/s00023-022-01166-0}}.

\bibitem{Nechita_2025}
Ion Nechita and Sang-Jun Park.
\newblock Random covariant quantum channels.
\newblock {\em Annales Henri Poincaré}, March 2025.
\newblock URL: \url{http://dx.doi.org/10.1007/s00023-025-01558-y}, \href {https://doi.org/10.1007/s00023-025-01558-y} {\path{doi:10.1007/s00023-025-01558-y}}.

\bibitem{Holevo2011}
Alexander~S. Holevo.
\newblock {\em Probabilistic and Statistical Aspects of Quantum Theory}.
\newblock Springer Science and Business Media, Berlin, 2011.
\newblock \href {https://doi.org/10.1007/978-88-7642-378-9} {\path{doi:10.1007/978-88-7642-378-9}}.

\bibitem{holevo2002}
Alexander~S. Holevo.
\newblock Remarks on the classical capacity of quantum channel, 2002.
\newblock URL: \url{https://arxiv.org/pdf/quant-ph/0212025.pdf}, \href {https://arxiv.org/abs/quant-ph/0212025} {\path{arXiv:quant-ph/0212025}}.

\bibitem{Jozsa1996}
Paul Hausladen, Richard Jozsa, Benjamin Schumacher, Michael Westmoreland, and William~K. Wootters.
\newblock Classical information capacity of a quantum channel.
\newblock {\em Phys. Rev. A}, 54:1869--1876, Sep 1996.
\newblock URL: \url{https://link.aps.org/doi/10.1103/PhysRevA.54.1869}, \href {https://doi.org/10.1103/PhysRevA.54.1869} {\path{doi:10.1103/PhysRevA.54.1869}}.

\bibitem{Datta2005}
Nilanjana Datta, Alexander~S. Holevo, and Y.M. Suhov.
\newblock On a sufficient condition for additivity in quantum information theory.
\newblock {\em Problems of Information Transmission}, 41:76, 2005.
\newblock \href {https://doi.org/10.1007/s11122-005-0013-7} {\path{doi:10.1007/s11122-005-0013-7}}.

\bibitem{Mozrzymas_2017}
Marek Mozrzymas, Michał Studziński, and Nilanjana Datta.
\newblock Structure of irreducibly covariant quantum channels for finite groups.
\newblock {\em Journal of Mathematical Physics}, 58(5), 2017.
\newblock URL: \url{http://dx.doi.org/10.1063/1.4983710}, \href {https://doi.org/10.1063/1.4983710} {\path{doi:10.1063/1.4983710}}.

\bibitem{Werner2002}
Reinhard~F. Werner and Alexander~S. Holevo.
\newblock Counterexample to an additivity conjecture for output purity of quantum channels.
\newblock {\em Journal of Mathematical Physics}, 43(9):4353–4357, September 2002.
\newblock URL: \url{http://dx.doi.org/10.1063/1.1498491}, \href {https://doi.org/10.1063/1.1498491} {\path{doi:10.1063/1.1498491}}.

\bibitem{Fannes2004}
M.~Fannes, B.~Haegeman, M.~Mosonyi, and D.~Vanpeteghem.
\newblock Additivity of minimal entropy output for a class of covariant channels, 2004.
\newblock URL: \url{https://arxiv.org/pdf/quant-ph/0410195}, \href {https://arxiv.org/abs/quant-ph/0410195} {\path{arXiv:quant-ph/0410195}}.

\bibitem{Holevo:2005}
Alexander~S. Holevo.
\newblock {Additivity Conjecture and Covariant Channels}.
\newblock {\em International Journal Quant. Information}, 03(01):41--47, 2005.
\newblock \href {https://doi.org/10.1142/s0219749905000530} {\path{doi:10.1142/s0219749905000530}}.

\bibitem{Fukuda2006}
Nilanjana Datta, Motohisa Fukuda, and Alexander~S. Holevo.
\newblock Complementarity and additivity for covariant channels.
\newblock {\em Quantum Inf. Process.}, 5:179, 2006.
\newblock \href {https://doi.org/10.1007/s11128-006-0021-6} {\path{doi:10.1007/s11128-006-0021-6}}.

\bibitem{Fukuda2017}
Motohisa Fukuda and Gilad Gour.
\newblock Additive bounds of minimum output entropies for unital channels and an exact qubit formula.
\newblock {\em IEEE Transactions on Information Theory}, 63:1818, 2017.
\newblock \href {https://doi.org/10.1109/TIT.2016.2641455} {\path{doi:10.1109/TIT.2016.2641455}}.

\bibitem{Faist_2020}
Philippe Faist, Sepehr Nezami, Victor~V. Albert, Grant Salton, Fernando Pastawski, Patrick Hayden, and John Preskill.
\newblock Continuous symmetries and approximate quantum error correction.
\newblock {\em Physical Review X}, 10(4), October 2020.
\newblock URL: \url{http://dx.doi.org/10.1103/PhysRevX.10.041018}, \href {https://doi.org/10.1103/physrevx.10.041018} {\path{doi:10.1103/physrevx.10.041018}}.

\bibitem{Zhou_2021}
Sisi Zhou, Zi-Wen Liu, and Liang Jiang.
\newblock New perspectives on covariant quantum error correction.
\newblock {\em Quantum}, 5:521, August 2021.
\newblock URL: \url{http://dx.doi.org/10.22331/q-2021-08-09-521}, \href {https://doi.org/10.22331/q-2021-08-09-521} {\path{doi:10.22331/q-2021-08-09-521}}.

\bibitem{Hayden2021}
Patrick Hayden, Sepehr Nezami, Sandu Popescu, and Grant Salton.
\newblock Error correction of quantum reference frame information.
\newblock {\em PRX Quantum}, 2:010326, Feb 2021.
\newblock URL: \url{https://link.aps.org/doi/10.1103/PRXQuantum.2.010326}, \href {https://doi.org/10.1103/PRXQuantum.2.010326} {\path{doi:10.1103/PRXQuantum.2.010326}}.

\bibitem{Gschwendtner_2021}
Martina Gschwendtner, Andreas Bluhm, and Andreas Winter.
\newblock Programmability of covariant quantum channels.
\newblock {\em Quantum}, 5:488, June 2021.
\newblock URL: \url{http://dx.doi.org/10.22331/q-2021-06-29-488}, \href {https://doi.org/10.22331/q-2021-06-29-488} {\path{doi:10.22331/q-2021-06-29-488}}.

\bibitem{Christandl2019}
Matthias Christandl, Alexander Müller-Hermes, and Michael~M. Wolf.
\newblock When do composed maps become entanglement breaking?
\newblock {\em Annales Henri Poincaré}, 20(7):2295–2322, February 2019.
\newblock URL: \url{http://dx.doi.org/10.1007/s00023-019-00774-7}, \href {https://doi.org/10.1007/s00023-019-00774-7} {\path{doi:10.1007/s00023-019-00774-7}}.

\bibitem{bauml2019}
Stefan Bäuml, Siddhartha Das, Xin Wang, and Mark~M. Wilde.
\newblock Resource theory of entanglement for bipartite quantum channels, 2019.
\newblock URL: \url{https://arxiv.org/abs/1907.04181}, \href {https://arxiv.org/abs/1907.04181} {\path{arXiv:1907.04181}}.

\bibitem{Bardet_2020}
Ivan Bardet, Benoît Collins, and Gunjan Sapra.
\newblock Characterization of equivariant maps and application to entanglement detection.
\newblock {\em Annales Henri Poincaré}, 21(10):3385–3406, August 2020.
\newblock URL: \url{http://dx.doi.org/10.1007/s00023-020-00941-1}, \href {https://doi.org/10.1007/s00023-020-00941-1} {\path{doi:10.1007/s00023-020-00941-1}}.

\bibitem{datta2004}
Nilanjana Datta, Alexander~S. Holevo, and Yuri Suhov.
\newblock Additivity for transpose depolarizing channels, 2004.
\newblock URL: \url{https://arxiv.org/abs/quant-ph/0412034}, \href {https://arxiv.org/abs/quant-ph/0412034} {\path{arXiv:quant-ph/0412034}}.

\bibitem{Pirandola2011}
Stefano Pirandola, Cosmo Lupo, Vittorio Giovannetti, Stefano Mancini, and Samuel~L Braunstein.
\newblock Quantum reading capacity.
\newblock {\em New Journal of Physics}, 13(11):113012, November 2011.
\newblock URL: \url{http://dx.doi.org/10.1088/1367-2630/13/11/113012}, \href {https://doi.org/10.1088/1367-2630/13/11/113012} {\path{doi:10.1088/1367-2630/13/11/113012}}.

\bibitem{Pirandola2017}
Stefano Pirandola, Riccardo Laurenza, Carlo Ottaviani, and Leonardo Banchi.
\newblock Fundamental limits of repeaterless quantum communications.
\newblock {\em Nature Communications}, 8(1), April 2017.
\newblock URL: \url{http://dx.doi.org/10.1038/ncomms15043}, \href {https://doi.org/10.1038/ncomms15043} {\path{doi:10.1038/ncomms15043}}.

\bibitem{Wilde_2017}
Mark~M. Wilde, Marco Tomamichel, and Mario Berta.
\newblock Converse bounds for private communication over quantum channels.
\newblock {\em IEEE Transactions on Information Theory}, 63(3):1792–1817, March 2017.
\newblock URL: \url{http://dx.doi.org/10.1109/TIT.2017.2648825}, \href {https://doi.org/10.1109/tit.2017.2648825} {\path{doi:10.1109/tit.2017.2648825}}.

\bibitem{Das2019}
Siddhartha Das and Mark~M. Wilde.
\newblock Quantum reading capacity: General definition and bounds.
\newblock {\em IEEE Transactions on Information Theory}, 65(11):7566–7583, November 2019.
\newblock URL: \url{http://dx.doi.org/10.1109/TIT.2019.2929925}, \href {https://doi.org/10.1109/tit.2019.2929925} {\path{doi:10.1109/tit.2019.2929925}}.

\bibitem{Das_2021}
Siddhartha Das, Stefan Bäuml, Marek Winczewski, and Karol Horodecki.
\newblock Universal limitations on quantum key distribution over a network.
\newblock {\em Physical Review X}, 11(4), October 2021.
\newblock URL: \url{http://dx.doi.org/10.1103/PhysRevX.11.041016}, \href {https://doi.org/10.1103/physrevx.11.041016} {\path{doi:10.1103/physrevx.11.041016}}.

\bibitem{sohail2025}
Sohail, Vivek Pandey, Uttam Singh, and Siddhartha Das.
\newblock Fundamental limitations on the recoverability of quantum processes.
\newblock {\em Annales Henri Poincar{\'e}}, July 2025.
\newblock \href {https://doi.org/10.1007/s00023-025-01590-y} {\path{doi:10.1007/s00023-025-01590-y}}.

\bibitem{AlNuwairan2014}
Muneerah~Al Nuwairan.
\newblock The extreme points of su(2)-irreducibly covariant channels.
\newblock {\em International Journal of Mathematics}, 25(6):1450048, 2014.
\newblock \href {https://doi.org/10.1142/S0129167X14500487} {\path{doi:10.1142/S0129167X14500487}}.

\bibitem{Lee2022}
Hun~Hee Lee and Sang-Gyun Youn.
\newblock Quantum channels with quantum group symmetry.
\newblock {\em Communications in Mathematical Physics}, 389(3):1303--1329, 2022.
\newblock \href {https://doi.org/10.1007/s00220-021-04283-9} {\path{doi:10.1007/s00220-021-04283-9}}.

\bibitem{Memarzadeh2022}
Laleh Memarzadeh and Barry~C. Sanders.
\newblock Group-covariant extreme and quasiextreme channels.
\newblock {\em Physical Review Research}, 4(3), September 2022.
\newblock URL: \url{http://dx.doi.org/10.1103/PhysRevResearch.4.033206}, \href {https://doi.org/10.1103/physrevresearch.4.033206} {\path{doi:10.1103/physrevresearch.4.033206}}.

\bibitem{Gour2021}
Gilad Gour and Mark~M. Wilde.
\newblock Entropy of a quantum channel.
\newblock {\em Physical Review Research}, 3:023096, May 2021.
\newblock URL: \url{https://link.aps.org/doi/10.1103/PhysRevResearch.3.023096}, \href {https://doi.org/10.1103/PhysRevResearch.3.023096} {\path{doi:10.1103/PhysRevResearch.3.023096}}.

\bibitem{Felix2018}
Felix Leditzky, Eneet Kaur, Nilanjana Datta, and Mark~M. Wilde.
\newblock Approaches for approximate additivity of the {H}olevo information of quantum channels.
\newblock {\em Physical Review A}, 97:012332, January 2018.
\newblock URL: \url{https://journals.aps.org/pra/abstract/10.1103/PhysRevA.97.012332}.

\bibitem{Yuan2019}
Xiao Yuan.
\newblock Hypothesis testing and entropies of quantum channels.
\newblock {\em Physical Review A}, 99:032317, March 2019.
\newblock URL: \url{https://link.aps.org/doi/10.1103/PhysRevA.99.032317}, \href {https://doi.org/10.1103/PhysRevA.99.032317} {\path{doi:10.1103/PhysRevA.99.032317}}.

\bibitem{Chiribella}
Giulio Chiribella, Giacomo~M. DAriano, and Paolo Perinotti.
\newblock Transforming quantum operations: Quantum supermaps.
\newblock {\em Europhysics Letters}, 83(3):30004, July 2008.
\newblock URL: \url{https://dx.doi.org/10.1209/0295-5075/83/30004}, \href {https://doi.org/10.1209/0295-5075/83/30004} {\path{doi:10.1209/0295-5075/83/30004}}.

\bibitem{Giulio2008a}
Giulio Chiribella, Giacomo~M. DAriano, and Paolo Perinotti.
\newblock Quantum circuit architecture.
\newblock {\em Physical Review Letters}, 101:060401, August 2008.
\newblock URL: \url{https://link.aps.org/doi/10.1103/PhysRevLett.101.060401}, \href {https://doi.org/10.1103/PhysRevLett.101.060401} {\path{doi:10.1103/PhysRevLett.101.060401}}.

\bibitem{Zyczkowski2008}
Karol \.Zyczkowski.
\newblock Quartic quantum theory: an extension of the standard quantum mechanics.
\newblock {\em Journal of Physics A: Mathematical and Theoretical}, 41(35):355302, July 2008.
\newblock URL: \url{http://dx.doi.org/10.1088/1751-8113/41/35/355302}, \href {https://doi.org/10.1088/1751-8113/41/35/355302} {\path{doi:10.1088/1751-8113/41/35/355302}}.

\bibitem{Giulio2013}
Giulio Chiribella, Alessandro Toigo, and Veronica Umanità.
\newblock Normal completely positive maps on the space of quantum operations.
\newblock {\em Open Systems and Information Dynamics}, 20(01):1350003, March 2013.
\newblock URL: \url{http://dx.doi.org/10.1142/S1230161213500030}, \href {https://doi.org/10.1142/s1230161213500030} {\path{doi:10.1142/s1230161213500030}}.

\bibitem{Gour_2019}
Gilad Gour.
\newblock Comparison of quantum channels by superchannels.
\newblock {\em {IEEE} Transactions on Information Theory}, 65(9):5880--5904, September 2019.
\newblock URL: \url{https://doi.org/10.1109%2Ftit.2019.2907989}, \href {https://doi.org/10.1109/tit.2019.2907989} {\path{doi:10.1109/tit.2019.2907989}}.

\bibitem{Giulio2008}
Giulio Chiribella, Giacomo~M. DAriano, and Paolo Perinotti.
\newblock Memory effects in quantum channel discrimination.
\newblock {\em Physical Review Letters}, 101(18), October 2008.
\newblock URL: \url{http://dx.doi.org/10.1103/PhysRevLett.101.180501}, \href {https://doi.org/10.1103/physrevlett.101.180501} {\path{doi:10.1103/physrevlett.101.180501}}.

\bibitem{Giulio2008b}
Giulio Chiribella, Giacomo~Mauro DAriano, and Paolo Perinotti.
\newblock Optimal cloning of unitary transformation.
\newblock {\em Physical Review Letters}, 101:180504, October 2008.
\newblock URL: \url{https://link.aps.org/doi/10.1103/PhysRevLett.101.180504}, \href {https://doi.org/10.1103/PhysRevLett.101.180504} {\path{doi:10.1103/PhysRevLett.101.180504}}.

\bibitem{Ziman2008}
Mário Ziman.
\newblock Process positive-operator-valued measure: A mathematical framework for the description of process tomography experiments.
\newblock {\em Physical Review A}, 77(6), June 2008.
\newblock URL: \url{http://dx.doi.org/10.1103/PhysRevA.77.062112}, \href {https://doi.org/10.1103/physreva.77.062112} {\path{doi:10.1103/physreva.77.062112}}.

\bibitem{Bisio2009}
A.~Bisio, Giulio Chiribella, Giacomo~M. DAriano, S.~Facchini, and Paolo Perinotti.
\newblock Optimal quantum tomography of states, measurements, and transformations.
\newblock {\em Physical Review Letters}, 102(1), January 2009.
\newblock URL: \url{http://dx.doi.org/10.1103/PhysRevLett.102.010404}, \href {https://doi.org/10.1103/physrevlett.102.010404} {\path{doi:10.1103/physrevlett.102.010404}}.

\bibitem{Giulio2009}
Giulio Chiribella, Giacomo~Mauro D’Ariano, and Paolo Perinotti.
\newblock Optimal covariant quantum networks.
\newblock In {\em AIP Conference Proceedings}, page 47–56. AIP, 2009.
\newblock URL: \url{http://dx.doi.org/10.1063/1.3131375}, \href {https://doi.org/10.1063/1.3131375} {\path{doi:10.1063/1.3131375}}.

\bibitem{Giulio2009a}
Giulio Chiribella, Giacomo~Mauro D'Ariano, and Paolo Perinotti.
\newblock Theoretical framework for quantum networks.
\newblock {\em Physical Review A}, 80:022339, August 2009.
\newblock URL: \url{https://link.aps.org/doi/10.1103/PhysRevA.80.022339}, \href {https://doi.org/10.1103/PhysRevA.80.022339} {\path{doi:10.1103/PhysRevA.80.022339}}.

\bibitem{Bisio2010}
Alessandro Bisio, Giulio Chiribella, Giacomo~Mauro D'Ariano, and Paolo Perinotti.
\newblock Information-disturbance tradeoff in estimating a unitary transformation.
\newblock {\em Physical Review A}, 82:062305, Dec 2010.
\newblock URL: \url{https://link.aps.org/doi/10.1103/PhysRevA.82.062305}, \href {https://doi.org/10.1103/PhysRevA.82.062305} {\path{doi:10.1103/PhysRevA.82.062305}}.

\bibitem{Wang_2023}
Xin Wang and Mark~M. Wilde.
\newblock Exact entanglement cost of quantum states and channels under positive-partial-transpose-preserving operations.
\newblock {\em Physical Review A}, 107(1), January 2023.
\newblock URL: \url{http://dx.doi.org/10.1103/PhysRevA.107.012429}, \href {https://doi.org/10.1103/physreva.107.012429} {\path{doi:10.1103/physreva.107.012429}}.

\bibitem{Diaz2018}
María~García Díaz, Kun Fang, Xin Wang, Matteo Rosati, Michalis Skotiniotis, John Calsamiglia, and Andreas Winter.
\newblock Using and reusing coherence to realize quantum processes.
\newblock {\em Quantum}, 2:100, October 2018.
\newblock URL: \url{http://dx.doi.org/10.22331/q-2018-10-19-100}, \href {https://doi.org/10.22331/q-2018-10-19-100} {\path{doi:10.22331/q-2018-10-19-100}}.

\bibitem{Gour2021a}
Gilad Gour and Carlo~Maria Scandolo.
\newblock Entanglement of a bipartite channel.
\newblock {\em Physical Review A}, 103(6), June 2021.
\newblock URL: \url{http://dx.doi.org/10.1103/PhysRevA.103.062422}, \href {https://doi.org/10.1103/physreva.103.062422} {\path{doi:10.1103/physreva.103.062422}}.

\bibitem{Burniston2020}
John Burniston, Michael Grabowecky, Carlo~Maria Scandolo, Giulio Chiribella, and Gilad Gour.
\newblock Necessary and sufficient conditions on measurements of quantum channels.
\newblock {\em Proceedings of the Royal Society A: Mathematical, Physical and Engineering Sciences}, 476(2236), April 2020.
\newblock URL: \url{http://dx.doi.org/10.1098/rspa.2019.0832}, \href {https://doi.org/10.1098/rspa.2019.0832} {\path{doi:10.1098/rspa.2019.0832}}.

\bibitem{Pollock2018}
Felix~A. Pollock, C{\'e}sar Rodr{\'i}guez-Rosario, Thomas Frauenheim, Mauro Paternostro, and Kavan Modi.
\newblock Operational {M}arkov condition for quantum processes.
\newblock {\em Physical Review Letters}, 120:040405, Jan 2018.
\newblock URL: \url{https://link.aps.org/doi/10.1103/PhysRevLett.120.040405}, \href {https://doi.org/10.1103/PhysRevLett.120.040405} {\path{doi:10.1103/PhysRevLett.120.040405}}.

\bibitem{Pollock2018NonMarkovian}
Felix~A. Pollock, C{\'e}sar Rodr{\'i}guez-Rosario, Thomas Frauenheim, Mauro Paternostro, and Kavan Modi.
\newblock Non-{M}arkovian quantum processes: {C}omplete framework and efficient characterization.
\newblock {\em Physical Review A}, 97:012127, Jan 2018.
\newblock URL: \url{https://link.aps.org/doi/10.1103/PhysRevA.97.012127}, \href {https://doi.org/10.1103/PhysRevA.97.012127} {\path{doi:10.1103/PhysRevA.97.012127}}.

\bibitem{Milz2021Quantum}
Simon Milz and Kavan Modi.
\newblock Quantum stochastic processes and quantum non-markovian phenomena.
\newblock {\em PRX Quantum}, 2:030201, Jul 2021.
\newblock URL: \url{https://link.aps.org/doi/10.1103/PRXQuantum.2.030201}, \href {https://doi.org/10.1103/PRXQuantum.2.030201} {\path{doi:10.1103/PRXQuantum.2.030201}}.

\bibitem{Jencova2024Structure}
Anna Jen{\v{c}}ov{\'a}.
\newblock On the structure of higher order quantum maps, 2024.
\newblock URL: \url{https://arxiv.org/abs/2411.09256}, \href {https://arxiv.org/abs/2411.09256} {\path{arXiv:2411.09256}}.

\bibitem{Milz2024Characterising}
Simon Milz and Marco~T{\'u}lio Quintino.
\newblock Characterising transformations between quantum objects, `completeness' of quantum properties, and transformations without a fixed causal order.
\newblock {\em {Quantum}}, 8:1415, Jul 2024.
\newblock \href {https://doi.org/10.22331/q-2024-07-17-1415} {\path{doi:10.22331/q-2024-07-17-1415}}.

\bibitem{paulsen2002}
Vern Paulsen.
\newblock {\em Completely bounded maps and operator algebras}, volume~78.
\newblock Cambridge University Press, 2002.

\bibitem{book}
Erling St{\o}rmer.
\newblock {\em Positive Linear Maps of Operator Algebras}.
\newblock Springer, Berlin, January 2013.
\newblock \href {https://doi.org/10.1007/978-3-642-34369-8} {\path{doi:10.1007/978-3-642-34369-8}}.

\bibitem{Bhatia2006Positive}
Rajendra Bhatia.
\newblock {\em Positive Definite Matrices}.
\newblock Princeton Series in Applied Mathematics. Princeton University Press, Princeton, NJ, 2006.

\bibitem{Watrous2018Theory}
John Watrous.
\newblock {\em The Theory of Quantum Information}.
\newblock Cambridge University Press, Cambridge, 2018.

\bibitem{Wolf2012Quantum}
Michael~M. Wolf.
\newblock Quantum channels {\&} operations: {A} guided tour.
\newblock Lecture notes, 2012.
\newblock URL: \url{https://www-m5.ma.tum.de/foswiki/pub/M5/Allgemeines/MichaelWolf/QChannelLecture.pdf}.

\bibitem{CHOI1975285}
Man-Duen Choi.
\newblock Completely positive linear maps on complex matrices.
\newblock {\em Linear Algebra and its Applications}, 10(3):285 -- 290, 1975.
\newblock URL: \url{http://www.sciencedirect.com/science/article/pii/0024379575900750}, \href {https://doi.org/10.1016/0024-3795(75)90075-0} {\path{doi:10.1016/0024-3795(75)90075-0}}.

\bibitem{Jamiolkowski1972Linear}
Andrzej Jamio{\l}kowski.
\newblock Linear transformations which preserve trace and positive semidefiniteness of operators.
\newblock {\em Reports on Mathematical Physics}, 3(4):275--278, 1972.
\newblock \href {https://doi.org/10.1016/0034-4877(72)90011-0} {\path{doi:10.1016/0034-4877(72)90011-0}}.

\bibitem{dePillis1967Linear}
John de~Pillis.
\newblock Linear transformations which preserve {H}ermitian and positive semidefinite operators.
\newblock {\em Pacific Journal of Mathematics}, 23(1):129--137, 1967.
\newblock URL: \url{https://msp.org/pjm/1967/23-1/p14.xhtml}, \href {https://doi.org/10.2140/pjm.1967.23.129} {\path{doi:10.2140/pjm.1967.23.129}}.

\bibitem{Sudarshan1961Stochastic}
E.~C.~G. Sudarshan, P.~M. Mathews, and Jayaseetha Rau.
\newblock Stochastic dynamics of quantum-mechanical systems.
\newblock {\em Physical Review}, 121(3):920--924, Feb 1961.
\newblock \href {https://doi.org/10.1103/PhysRev.121.920} {\path{doi:10.1103/PhysRev.121.920}}.

\bibitem{Giulio2006}
Sergio De~Zordo.
\newblock {\em Quantum Supermaps}.
\newblock PhD thesis, Universit\`a degli Studi di Pavia, 2006.
\newblock URL: \url{https://wordpress.qubit.it/wp-content/uploads/thesis/tesidezordo.pdf}.

\bibitem{LandauStreater1993}
L.~J. Landau and R.~F. Streater.
\newblock On birkhoff's theorem for doubly stochastic completely positive maps of matrix algebras.
\newblock {\em Linear Algebra and its Applications}, 193:107--127, 1993.
\newblock \href {https://doi.org/10.1016/0024-3795(93)90274-R} {\path{doi:10.1016/0024-3795(93)90274-R}}.

\bibitem{Wolf2009}
C.~B. Mendl and M.~M. Wolf.
\newblock Unital quantum channels - convex structure and revivals of birkhoff's theorem.
\newblock {\em Communication in Mathematical Physics}, 289, 2009.
\newblock \href {https://doi.org/10.1007/s00220-009-0824-2} {\path{doi:10.1007/s00220-009-0824-2}}.

\bibitem{DC_AK_2006}
Dariusz Chru{\'s}ci{\'n}ski and Andrzej Kossakowski.
\newblock Class of positive partial transposition states.
\newblock {\em Physical Review A}, 74(2):022308, August 2006.
\newblock \href {https://doi.org/10.1103/PhysRevA.74.022308} {\path{doi:10.1103/PhysRevA.74.022308}}.

\bibitem{Chruscinski2022Dynamical}
Dariusz Chru{\'s}ci{\'n}ski.
\newblock Dynamical maps beyond {M}arkovian regime.
\newblock {\em Physics Reports}, 992:1--85, Dec 2022.
\newblock \href {https://doi.org/10.1016/j.physrep.2022.09.003} {\path{doi:10.1016/j.physrep.2022.09.003}}.

\bibitem{DC_Topical}
Dariusz Chru{\'s}ci{\'n}ski and Gniewomir Sarbicki.
\newblock Entanglement witnesses: construction, analysis and classification.
\newblock {\em Journal of Physics A: Mathematical and Theoretical}, 47(48):483001, November 2014.
\newblock \href {https://doi.org/10.1088/1751-8113/47/48/483001} {\path{doi:10.1088/1751-8113/47/48/483001}}.

\bibitem{DC_AK_OSID}
Dariusz Chru{\'s}ci{\'n}ski and Andrzej Kossakowski.
\newblock On the structure of entanglement witnesses and new class of positive indecomposable maps.
\newblock {\em Open Systems \& Information Dynamics}, 14(03):275--294, 2007.
\newblock \href {https://doi.org/10.1007/s11080-007-9052-4} {\path{doi:10.1007/s11080-007-9052-4}}.

\bibitem{circulant_states}
Dariusz Chru\ifmmode \acute{s}\else \'{s}\fi{}ci\ifmmode~\acute{n}\else \'{n}\fi{}ski and Andrzej Kossakowski.
\newblock Circulant states with positive partial transpose.
\newblock {\em Phys. Rev. A}, 76:032308, Sep 2007.
\newblock URL: \url{https://link.aps.org/doi/10.1103/PhysRevA.76.032308}, \href {https://doi.org/10.1103/PhysRevA.76.032308} {\path{doi:10.1103/PhysRevA.76.032308}}.

\bibitem{Korzekwa2018}
Kamil Korzekwa, Stanislaw Czach\'orski, Zbigniew Puchala, and Karol \.Zyczkowski.
\newblock Coherifying quantum channels.
\newblock {\em New Journal of Physics}, 20(4):043028, apr 2018.
\newblock \href {https://doi.org/10.1088/1367-2630/aaaff3} {\path{doi:10.1088/1367-2630/aaaff3}}.

\bibitem{Puchala2021}
Zbigniew Puchala, Kamil Korzekwa, Roberto Salazar, Pawel Horodecki, and Karol \.Zyczkowski.
\newblock Dephasing supechannels.
\newblock {\em Physical Review A}, 104:052611, Nov 2021.
\newblock URL: \url{https://link.aps.org/doi/10.1103/PhysRevA.104.052611}, \href {https://doi.org/10.1103/PhysRevA.104.052611} {\path{doi:10.1103/PhysRevA.104.052611}}.

\bibitem{Rico2025}
Albert Rico, Moisés~Bermejo Morán, Fereshte Shahbeigi, and Karol Życzkowski.
\newblock Certifying nonlocal properties of noisy quantum operations.
\newblock {\em Quantum}, 9:1807, 2025.
\newblock \href {https://doi.org/10.22331/q-2025-07-22-1807} {\path{doi:10.22331/q-2025-07-22-1807}}.

\bibitem{Bihalan}
D.~Chru\'sci\'nski and B.~Bhattacharya.
\newblock A class of {S}chwarz qubit maps with diagonal unitary and orthogonal symmetries.
\newblock {\em J. Phys. A: Math. Theor.}, 57:395202, 2024.
\newblock \href {https://doi.org/10.1088/1751-8121/ad75d6} {\path{doi:10.1088/1751-8121/ad75d6}}.

\bibitem{Christandl2012}
Christandl Matthias.
\newblock {PPT} square conjecture.
\newblock {\em Banff International Research Station Workshop: Operator Structures in Quantum Information Theory}, 2012.

\bibitem{Jencova2026Order}
Anna Jen{\v{c}}ov{\'a}.
\newblock Order structure and signalling in higher order quantum maps, 2026.
\newblock URL: \url{https://arxiv.org/abs/2604.09192}, \href {https://arxiv.org/abs/2604.09192} {\path{arXiv:2604.09192}}.

\bibitem{Zambon2024Process}
Guilherme Zambon.
\newblock Process tensor distinguishability measures.
\newblock {\em Physical Review A}, 110:042210, Oct 2024.
\newblock URL: \url{https://link.aps.org/doi/10.1103/PhysRevA.110.042210}, \href {https://doi.org/10.1103/PhysRevA.110.042210} {\path{doi:10.1103/PhysRevA.110.042210}}.

\end{thebibliography}
\end{document}